\documentclass{article}
\usepackage{epsfig}
\usepackage{amsmath}
\def\be{\begin{equation}}
\def\ee{\end{equation}}
\def\bem{\begin{multline}}
\def\eem{\end{multline}}
\def\bea{\begin{eqnarray}}
\def\eea{\end{eqnarray}}

\def\xbzero{{\bf x}_0}
\def\xbone{{\bf x}_1}
\def\kbone{{\bf k}_1}
\def\xbtwo{{\bf x}_2}

\def\bb{{\bf b}}

\def\bzerone{{\bf b}_{01}}
\def\xzerone{{\bf x}_{01}}
\def\xonetwo{{\bf x}_{12}}
\def\xtwozer{{\bf x}_{20}}
\def\asb{{\bar{\alpha}_s}}

\long\def\symbolfootnote[#1]#2{\begingroup%
\def\thefootnote{\fnsymbol{footnote}}\footnote[#1]{#2}\endgroup}
\newcommand\prd[3]{{\it Phys. Rev. }{\bf D #1} (#2) #3} 
\newcommand\npb[3]{{\it Nucl. Phys. }{\bf B #1} (#2) #3}  
\newcommand\npa[3]{{\it Nucl. Phys. }{\bf A #1} (#2) #3} 
\newcommand\prl[3]{{\it Phys. Rev. Lett. }{\bf #1} (#2) #3} 
\newcommand\plb[3]{{\it Phys. Lett. }{\bf B #1} (#2) #3}
\def\gapproxeq{\lower .7ex\hbox{$\;\stackrel{\textstyle
>}{\sim}\;$}}

\begin{document}
\begin{flushright}
BNL-NT-04/37
\end{flushright}
\vspace*{1in}
\begin{center}
{\Large \bf Nonlinear evolution equations in QCD
\symbolfootnote[2]{Presented at XLIV Cracow School of Theoretical Physics, June 2004, Zakopane, Poland.}}\\
\vspace*{0.4in}
Anna M. Sta\'sto$^{(a,b)}$ \\
\vspace*{0.5cm}
$^{(a)}$ {\it Nuclear Theory Group, Physics Department, \\
Brookhaven National Laboratory,\\
 Upton, NY 11973, USA\symbolfootnote[1]{Permanent address.}}\\
and \\
$^{(b)}$
{\it H.~Niewodnicza\'nski Institute of Nuclear Physics, \\
Polish Academy of Sciences,\\
 ul. Radzikowskiego 152,  31-342 Krak\'ow, Poland}
\vskip 2mm  
\end{center}  
\vspace*{1cm}  
\centerline{(\today)}  
  
\vskip1cm 

\begin{abstract}
The following  lectures are an introduction to the phenomena  of partonic saturation and nonlinear evolution equations in Quantum Chromodynamics. After a short introduction to the linear evolution, the problems of unitarity bound and  parton saturation are discussed. The nonlinear Balitsky-Kovchegov evolution equation in the high energy limit is introduced, and
the progress towards the understanding of the properties of its solution is reviewed. We discuss the concepts of the saturation scale,  geometrical scaling and the lack of the infrared diffusion.  Finally, we give a brief summary of current theoretical developments
which extend beyond the Balitsky-Kovchegov equation.
\end{abstract}

\section{Introduction}
\begin{figure}
\centerline{\epsfig{file=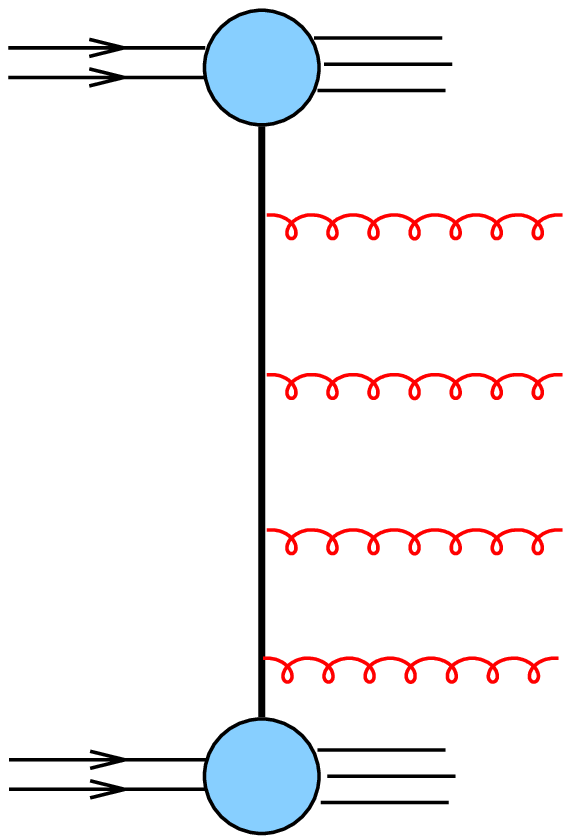,height=5cm}\hspace*{3cm}\epsfig{file=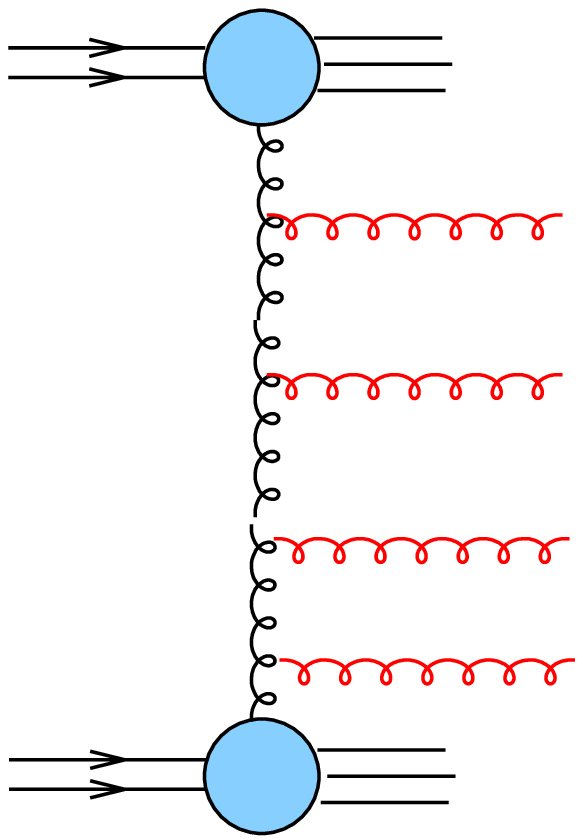,height=5cm}}
\caption{Scattering of two hadronic probes at high energy. Left: QED-type diagrams; right: diagrams with gluon self-interactions.}
\label{fig:1}
\end{figure}

One of the most intriguing problems of Quantum Chromodynamics is the growth of   cross sections for hadronic interactions at high energies. Let us consider the scattering of   two particles at very high energy   as shown  in Fig.\ref{fig:1}. As the  energy grows, so does the  probability of emission of soft particles. In the case of QED one has to consider a diagram of the type 
shown on the left hand side graph in Fig.~\ref{fig:1}.  In QCD one also encounters   these   diagrams,  but there are also additional diagrams of the type shown on the right hand side of Fig.~\ref{fig:1}. Since gluons are the carriers of the color charge and couple to each other, the increase of energy  causes a fast growth of the gluon density and, consequently, of the cross section. This increase  leads to the formation of a dense,  colored medium at very high energies.
 
In perturbative QCD the growth of the gluon density in the limit of high energy 
is governed by the BFKL Pomeron equation \cite{BFKL}. The solution to this  equation gives a very strong, power-like growth
of the gluon density, and  of the resulting cross section
$$f(x) \; \sim \;  x^{-\lambda}\; ,$$
where $x$ is the Bjorken variable (a fraction of target's longitudinal momentum  ) and $\lambda = \frac{4 \ln2 N_c}{\pi} \alpha_s$ is the intercept of the perturbative Pomeron in the leading logarithmic (LLx, $\ln 1/x \gg 1$) approximation.
In the pioneering paper \cite{GLR}, Gribov, Levin and Ryskin pointed out that the gluon recombination is important 
at high energies. It reduces the growth of the parton density by producing an effect  called the {\it perturbative partonic saturation}.
In \cite{GLR}, a new nonlinear evolution equation in double leading logarithmic approximation (DLLA, $\ln 1/x \ln Q^2 \gg 1$) for the gluon density has been postulated.
In addition to the gluon production it also takes into account the recombination effects
\be
Q^2 \, \frac{\partial^2 xG(x,Q^2)}{\partial \ln 1/x \partial Q^2 } \; = \; \frac{\alpha_s N_c}{\pi} xG(x,Q^2) \; - \; 
\frac{4 \alpha_s^2 N_c}{3 C_F R^2}\frac{1}{Q^2} [xG(x,Q^2)]^2 \; .
\label{eq:glr}
\ee
Note the negative sign in front of the nonlinear term  responsible for the gluon recombination. The strong growth generated by the linear  term is damped for large  gluon densities $xG(x,Q^2)$ ( of the order $1/\alpha_s$). \\
Partonic saturation plays also  an important role in the context of the unitarity bound. It is well known that the hadronic cross sections  obey the Froissart bound \cite{Froissart} which stems from the general assumptions of the analyticity and unitarity of the scattering amplitude. The Froissart bound implies that the total cross section
does not grow faster than the logarithm squared of the energy 
\be
\sigma_{tot} \; = \; \frac{\pi}{m_{\pi}^2} \, (\ln s)^2 \; ,
\ee
where $m_{\pi}$ is the scale of the range of the strong force.
It is generally believed, that the parton saturation  mechanism leads to the unitarization of the cross section at high energies. Unfortunately, the problem is quite complex since the parton saturation is  purely perturbative mechanism while the Froissart bound has been derived  from general principles and it refers to the QCD as to a complete theory of strong interactions
( including the  nonperturbative effects). 
The GLR postulate resulted in increased efforts
 to develop a 
theory able to  describe the saturation at high energies. 
One effective theory for  high density partonic systems at small $x$ is
 the Color Glass Condensate \cite{CGC} with the resulting JIMWLK evolution equations  \cite{JIMWLK,Weigert}.
Another approach has been developed by Balitsky \cite{Balitsky} who constructed an infinite hierarchy of coupled equations for correlators of Wilson lines. In the mean field approximation
the first equation of this theory decouples, and is equivalent to the Kovchegov equation  \cite{Kov} (derived independently in the dipole approach \cite{Mueller}).
The Kovchegov equation  is a nonlinear equation for the dipole scattering amplitude  valid in the leading  $\log 1/x$ approximation. The Balitsky-Kovchegov (BK) equation is perhaps the best
known equation that  includes the saturation effects and has a virtue that it can be relatively simply solved, at least numerically. 

The goal of the  following  lecture is  to introduce the reader into the phenomenon of the partonic saturation and  nonlinear evolution using BK equation as an example. We shall start with a brief review of the linear evolution in QCD. Then we shall recall the Froissart bound and the necessary conditions for its derivation. Then the properties of the solution of the BK equation
will be investigated with particular emphasis on the 
 the infrared diffusion, saturation scale and the geometrical scaling. We will continue with  an analysis of this equation in  general case in 4 dimensions, which takes into account the  dipole spatial distribution in the impact parameter space.
We will conclude with  a short outlook and the discussion of recent theoretical developments in the field.
\section{DIS kinematics and variables}
\begin{figure}
\centerline{\epsfig{file=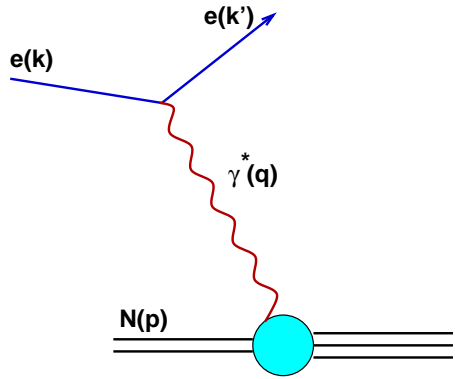,height=5cm}}
\caption{Deep inelastic scattering process of electron on the hadronic target.}
\label{fig:dis}
\end{figure}

Let us  concentrate on the deep inelastic scattering process of lepton off the hadron or nucleus. For completeness, let us  first recall  the basic kinematic features  of DIS, as represented in  Fig.~\ref{fig:dis}.
The total energy squared of the electron-nucleon system is $s_{(eN)}=(p+k)^2$, whereas that of the  photon-nucleon system is
$s_{\gamma^* N}= (p+q)^2$. The  photon virtuality is denoted by  $q^2=(k-k')^2=-Q^2<0$ and the Bjorken variable $x=\frac{Q^2}{2p\cdot q}=\frac{Q^2}{Q^2+s_{(\gamma ^* N)}}$.

The high energy regime, is defined  as
\begin{eqnarray}
& s_{(\gamma^*N)} & \longrightarrow  \infty \; , \nonumber \\
& x & = \; \frac{Q^2}{Q^2+s_{(\gamma ^* N)}} \;  \simeq \; \frac{Q^2}{s_{(\gamma ^* N)}}\nonumber \;  \longrightarrow \; 0 \; ,\\
& Y & =  \; \ln 1/x \; \longrightarrow \; \infty \; .
\end{eqnarray}
\section{The linear evolution equations of QCD}
\begin{figure}
\centerline{\epsfig{file=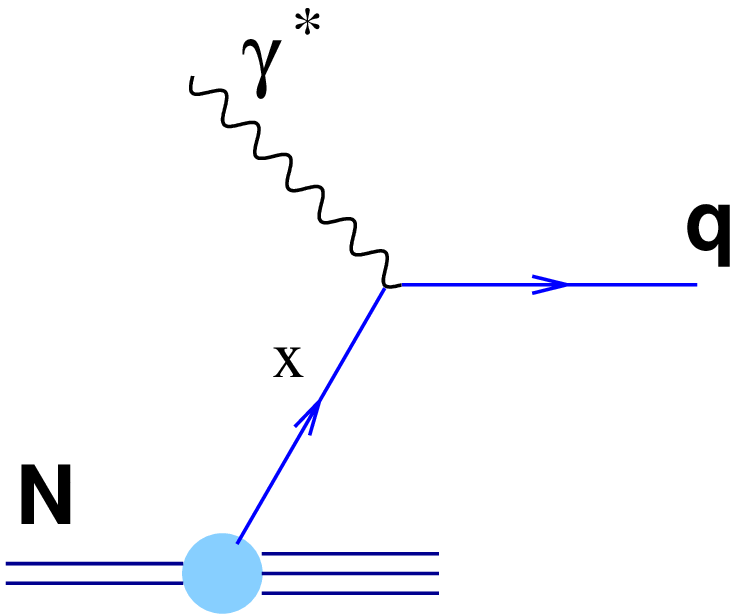,height=3cm}\hspace*{2cm}\epsfig{file=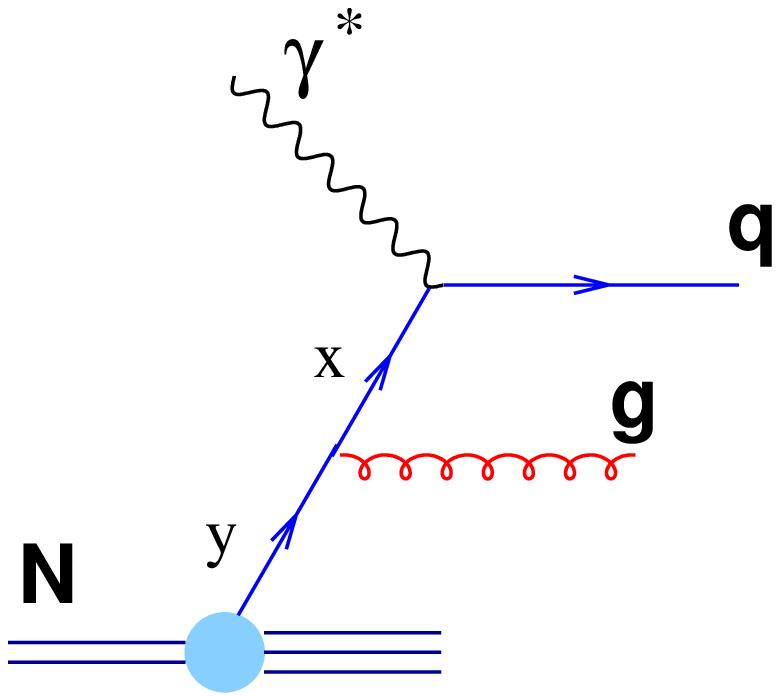,height=3cm}}
\caption{Left: photon with virtuality $Q^2$ probes quark with a longitudinal momentum fraction $x$. Right: the virtuality (resolution) is increased so the density of quarks also grows. }
\label{fig:splitqg}
\end{figure}

\begin{figure}
\centerline{\epsfig{file=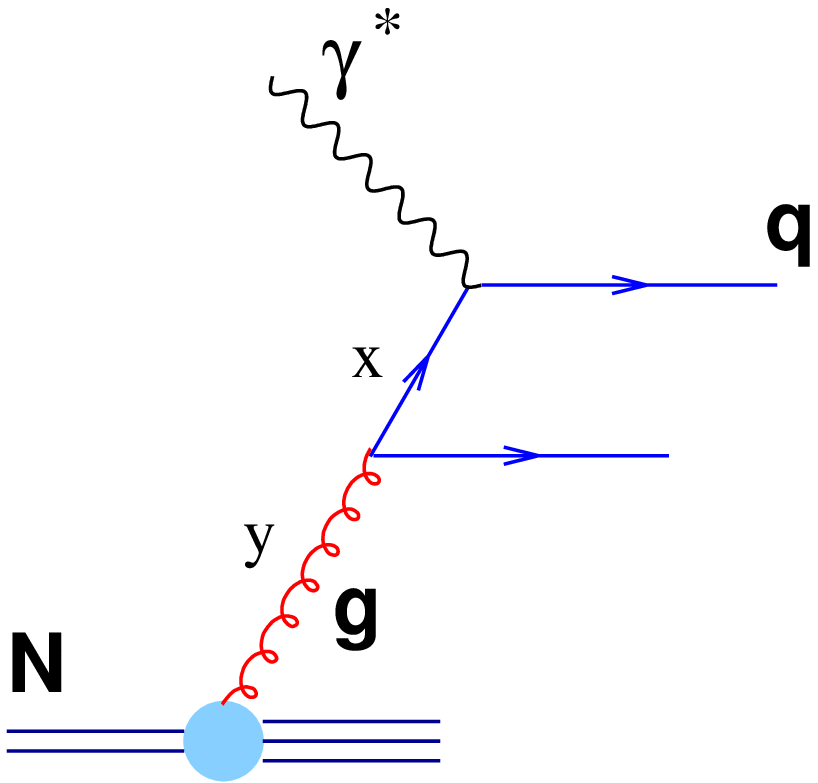,height=3cm}\hspace*{2cm}\epsfig{file=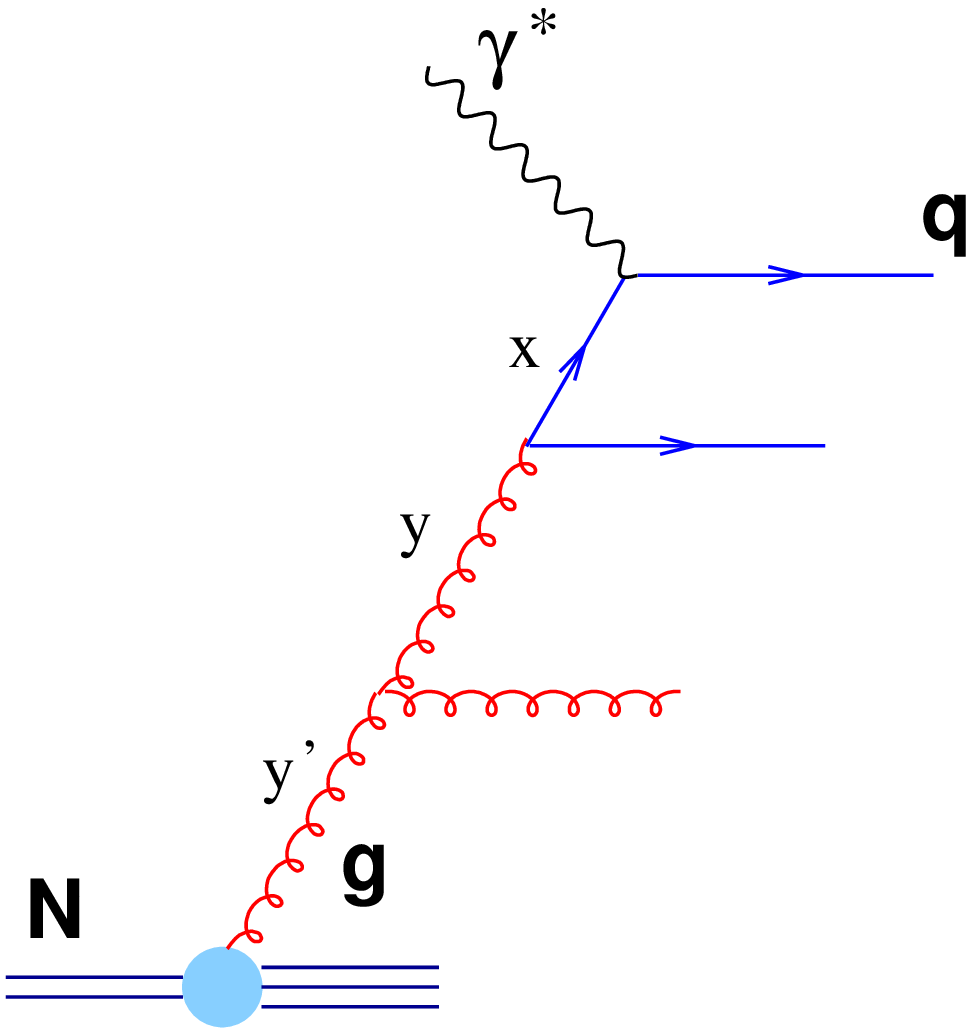,height=3.5cm}}
\caption{Additional set of diagrams present in the DGLAP evolution.}
\label{fig:splitqg1}
\end{figure}

Let us consider a scattering of photon with virtuality $Q^2$ off a hadron at center of mass energy $\sqrt{s}$.
The photon virtuality defines a resolution scale $\lambda \sim \frac{1}{\sqrt{Q^2}}$ with which one probes the partonic structure of a hadron, see the left hand side graph in Fig.~\ref{fig:splitqg}.
At a  given resolution $t=\ln Q^2/Q^2_0$, the  photon probes the density of partons $q(x,t)$ with a fraction of the  hadron momentum $x=Q^2/s$.
By increasing $Q^2$ one also increases  the resolution, and the density of quarks is larger: $q(x,t)+\delta q(x,t)$, see right-hand graph in Fig.~\ref{fig:splitqg}.
The underlying process can be described by the  following  linear evolution equation for density
\be
\frac{\partial q(x,t)}{\partial t} \; = \; \frac{\alpha_s(t)}{2\pi} \int_x^1 \frac{dy}{y} P_{qq}(x/y) q(y,t) \; .
\label{eq:dglap}
\ee
The splitting function $P_{qq}(z=x/y)$ describes the  probability of finding a  quark inside theparent quark, with a fraction $z=x/y$
of  the parent quark momentum.
This is  one of the set of the well known DGLAP evolution equations \cite{DGLAP}.
For the DGLAP equations to be complete, apart from the quark density $q(x,t)$, one has to include the gluon density $g(x,t)$  coupled to $q$.
The evolution of the gluon density is shown  in Fig.~\ref{fig:splitqg1}.
The full set of singlet DGLAP  equations reads as 
\begin{displaymath}
\frac{\partial}{\partial t} \,\left[
\begin{array}{c}
\Sigma(x,t) \\
g(x,t)
\end{array}
\right]
\;=\;
\frac{\alpha_s(t)}{2\pi}
\left[
\begin{array}{cc}
P_{qq} & 2N_f P_{qg} \\
P_{gq} &  P_{gg}
\end{array}
\right] \,
\stackrel{x}{\otimes}
\,
\left[
\begin{array}{c}
\Sigma(x,t) \\
g(x,t)
\end{array}
\right] \; ,
\nonumber
\end{displaymath}
where $\Sigma(x,t) = \sum_i [q_i(x,t)+\bar{q}_i(x,t)]$ is the singlet quark
density. \\

\begin{figure}[htb]
\centerline{\epsfig{file=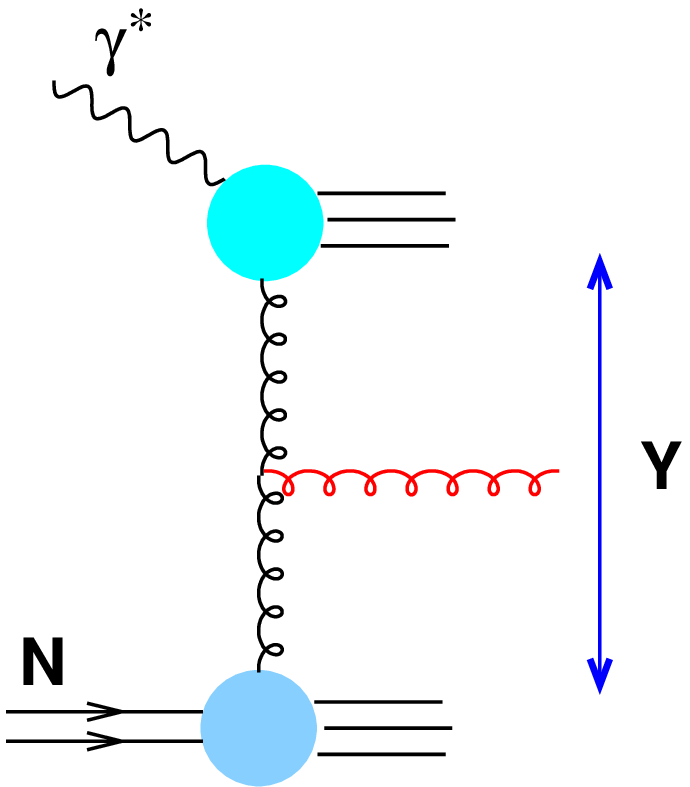,height=3cm}\hspace*{2cm}\epsfig{file=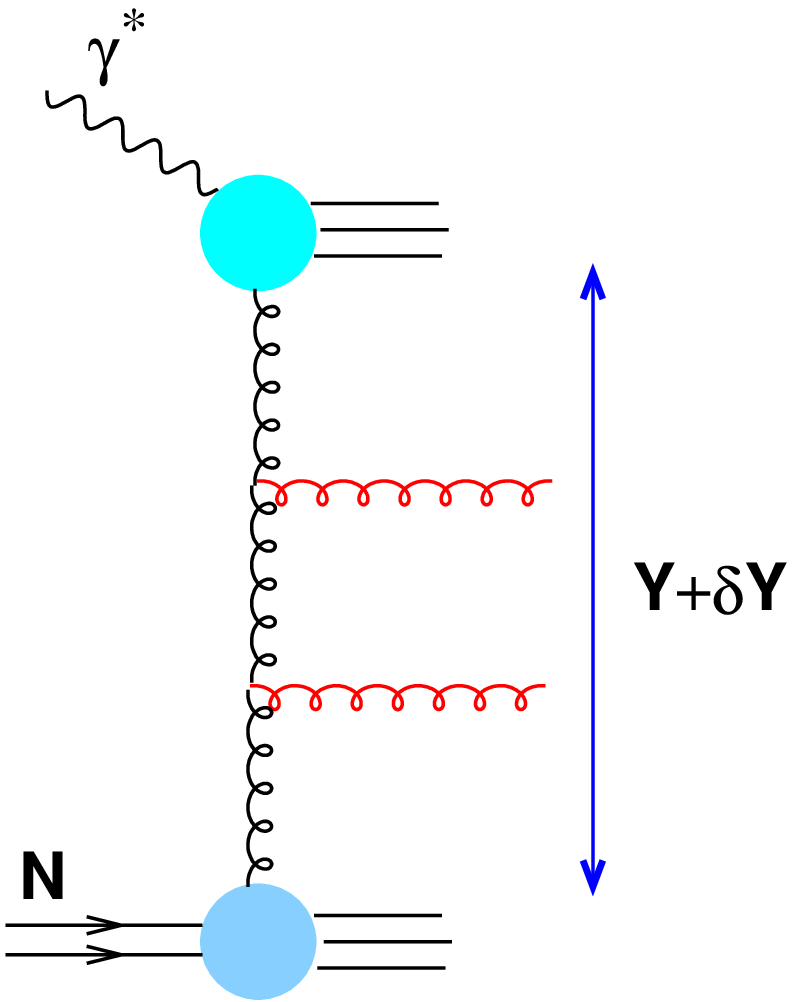,height=3.4cm}}
\caption{Evolution in the BFKL framework: virtualities of probes are fixed and energy is increased.}
\label{fig:bfklladder}
\end{figure}

An   alternative approach  to DGLAP \cite{BFKL} is to consider a fixed virtuality of the probe
and to increase the energy $s$ (or alternatively rapidity $Y$), see Fig.~\ref{fig:bfklladder}.
This procedure leads to the BFKL  equation
\be
\frac{\partial G(x,t)}{\partial \ln 1/x} \; = \; \frac{\alpha_s N_c}{\pi} \int dt' {\cal K}(t,t') G(x,t') \; ,
\label{eq:bfkl}
\ee
which is an evolution equation in Bjorken $x$. The quantity ${\cal K}(t,t')$ ( Lipatov kernel)  describes the probability of branching of the gluon with virtuality $t'$ into another gluon with of virtuality $t$.
The function $G(x,t)$ is called the unintegrated gluon density and  is related to $g(x,t)$  by 
$$
g(x,t) \; = \; \int^t \, dt' \, G(x,t') \; .
$$
Both ${\cal K}$ and $P_{ij}$  have perturbative expansions in $\alpha_s$ and share a finite number of common terms in the expansion.
The solution to the BFKL equation has the  form, see for example \cite{Lipatov86}
\be
G(x) \sim x^{-\lambda_P}\; , \hspace*{1cm} \lambda_P = 4 \ln 2 N_c \alpha_s/ \pi \, .
\label{eq:bfklsol}
\ee
This solution  strongly grows with $1/x$ (and, correspondingly with energy $s=Q^2/x$) which  contradicts the Froissart bound \cite{Froissart}.

\section{Froissart bound}
Let us  recall the basic assumptions used in  the derivation of the Froissart bound \cite{Froissart}.
This bound applies for the total cross section for scattering of two hadrons, and reads as follows
\be
\sigma_{\rm TOT} \; \le \frac{\pi}{m_{\pi}^2} (\ln s)^2 \; .
\ee
Its derivation is based on  two assumptions: the unitarity of partial amplitudes and 
the finite length of the strong interaction.
The first condition demands that the $S$ matrix  has to be unitary
\be
S^{\dagger} S \; = \; S S^{\dagger} \; = \; 1 \; .
\label{eq:s}
\ee
A set of particle states $|m\rangle$ has to satisfy completeness relation
\be
\sum_m |m \rangle \langle m|  \; = \; 1 \; .
\ee
Then the probability that the initial state $|i\rangle$ evolves into the final state $|f\rangle$ is
\be
P_{fi} = |\langle f | S | i \rangle |^2 \; .
\ee
Since a probability that a given final state comes from any initial state is 1 therefore 
the sum of these probabilities  over all the initial states equals 1 
\be
\sum_f P_{fi} \; = \; \langle i | S^{\dagger} S | i \rangle \; = \; 1 \; ,
\ee
which is interpreted as a unitarity condition for the S-matrix  (\ref{eq:s}).\\

The second  assumption is  on the finite range $R$ of strong force
determined by the mass scale $m_{\pi}$ 
\be
R \sim \frac{1}{m_{\pi}} \; ,
\ee
which determines the cutoff.
The scale $m_{\pi}$ is entirely nonperturbative.
The Froissart bound can be derived \cite{Froissart}
by using these two assumptions and the Mandelstam representation.
One has to stress that the Froissart bound  is applicable to the  complete QCD theory that includes both perturbative and non-perturbative
parts.

\begin{figure}[htb]
\centerline{\epsfig{file=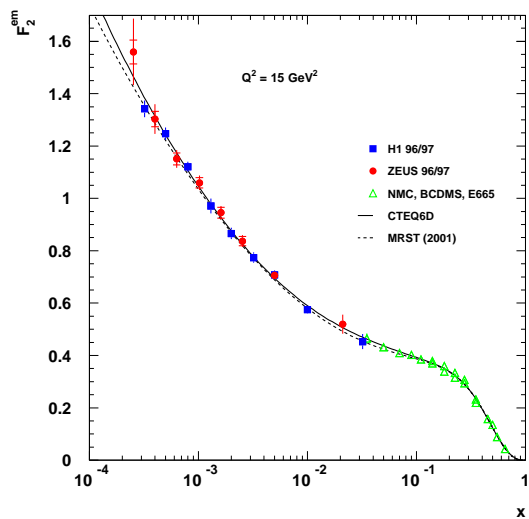,height=8cm}}
\caption{$F_2$ structure function data from HERA collider and fixed target experiments.}
\label{fig:F2}
\end{figure}

It is  worthwhile to mention that while one believes that Froissart bound is valid, it is not immediately visible  from the data. The data show that the structure function  for deep inelastic scattering at high photon virtualities strongly rises with  decreasing $x$,
consistent with a power behaviour $x^{-\lambda}$ with $\lambda\simeq 0.35$.
Further,  the data show no sign of a logarithmic dependence is seen, see Fig.~\ref{fig:F2}.

Pinning down the saturation effects in  experimental data is a  nontrivial task. A specific  problem with the structure function data is that it involves completely inclusive measurements. In particular, structure function $F_2$ is averaged over the impact parameter of the $\gamma^* - p$ collision.  It  is also  known that saturation effects crucially depend on this variable. Therefore the search for the saturation should involve  more exclusive processes, e.g. diffraction. For more information on the saturation phenomenology  consult \cite{MSM,GBW} and references therein.

\section{Parton saturation and nonlinear evolution}

\begin{figure}[htb]
\centerline{\epsfig{file=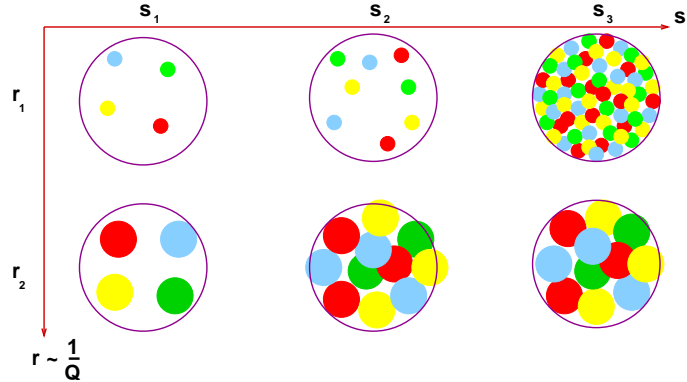,height=5cm}}
\caption{Schematic view of parton saturation. Horizontal axis is energy squared $s$, the vertical axis is $r$, the parton size.}
\label{fig:Sat}
\end{figure}

Given  that the Froissart bound   should be satisfied,
a natural question  arises: how to modify the perturbative evolution in order to tame the growth 
of the cross section? One would like to identify  the Feynman diagrams which are responsible for  gluon recombination and to derive the appropriate evolution equations that  include these  diagrams.
As already stated in the introduction, the standard evolution equations lead to a strong rise of the parton density in high energy limits.
One can expect that partons overlap when their density becomes large.
The schematic picture of the saturation phenomenon is shown in Fig.~\ref{fig:Sat}.

The horizontal axis represents the energy whereas and  the vertical one  the parton size defined by the inverse of the photon virtuality $r\sim 1/Q$ in the deep inelastic scattering process. The onset of saturation depends on energy and  on the parton size. The larger the size of the partons,
the earlier they  fill up the available area and start to reinteract. A further increase of the energy will not increase the  probed parton density.
Therefore, apart from  production diagrams, one has to include additional diagrams which take into account gluon recombination.
This leads to the modification of the evolution equation by a term which is nonlinear in  density. The first equation of this type 
was the GLR equation \cite{GLR} 
\be
Q^2 \, \frac{\partial^2 xG(x,Q^2)}{\partial \ln 1/x \partial Q^2 } \; = \; \frac{\alpha_s N_c}{\pi} xG(x,Q^2) \; - \; 
\frac{4 \alpha_s^2 N_c}{3 C_F R^2}\frac{1}{Q^2} [xG(x,Q^2)]^2 \; .
\label{eq:glr1}
\ee
The first term  in (\ref{eq:glr1}) is the usual DGLAP term in the double logarithmic approximation, $\ln 1/x \ln Q^2/\Lambda^2 \gg 1$, whereas the second, nonlinear, term is responsible for  gluon recombination. The nonlinear term is inversely proportional to the hadron area $\sim R^2$ and the scale $\sim Q^2$ at which the gluon density is  probed. The smaller the hadron area, the earlier the partons fill it up and saturate.
The scale $Q^2$ defines the parton size $r\sim 1/Q$. For small values of  $r$     the saturation is delayed to larger energies.

\begin{figure}[htb]
\centerline{\epsfig{file=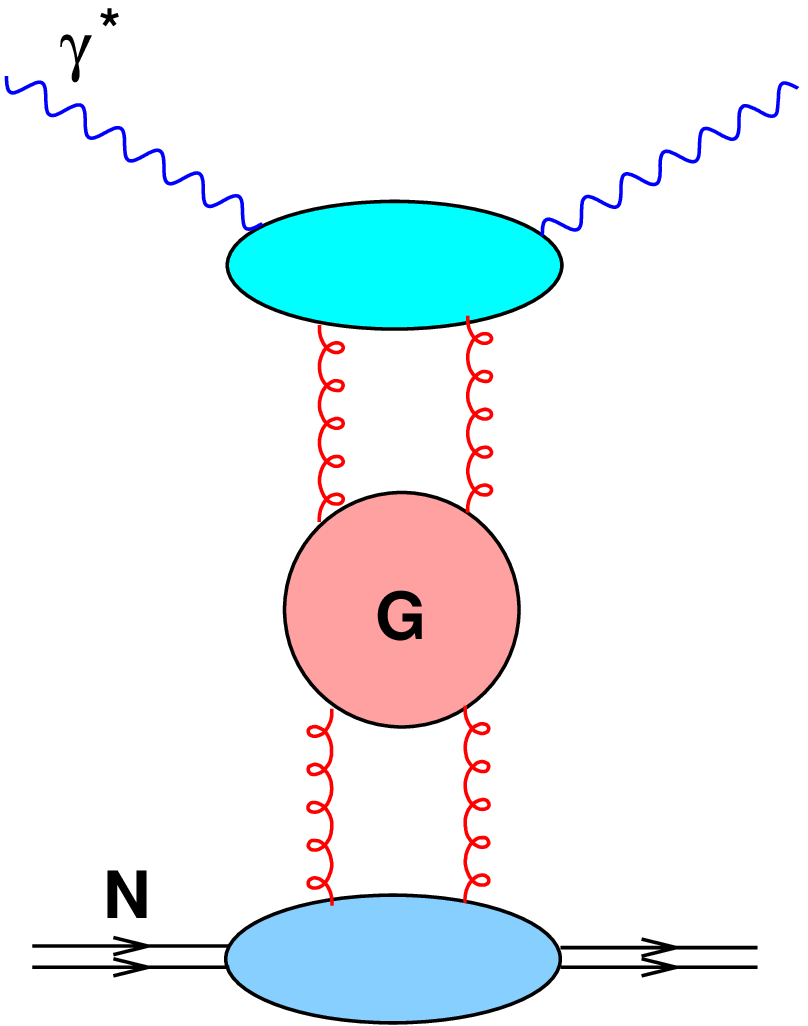,height=6cm}\hfill\epsfig{file=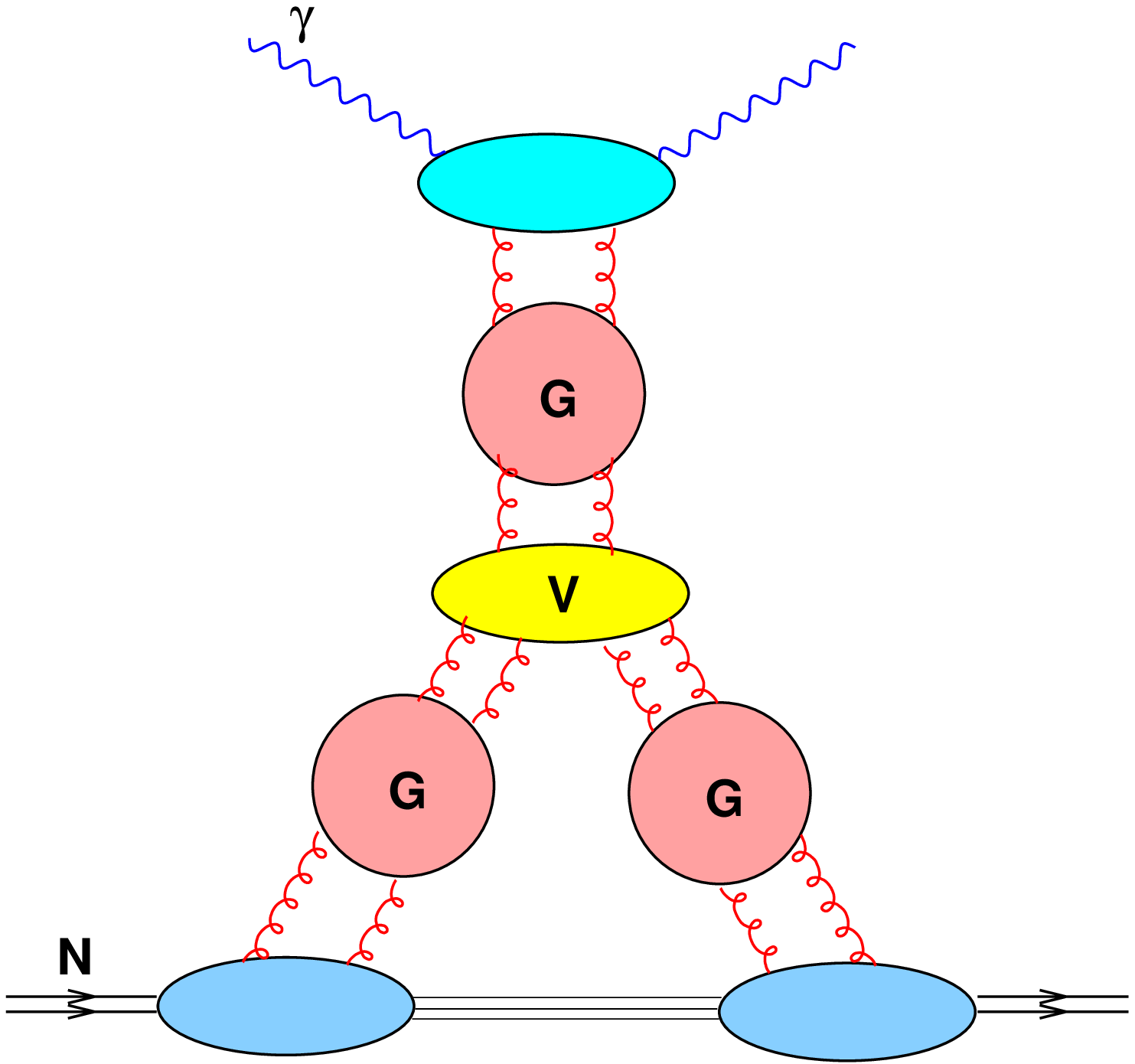,height=6cm}}
\caption{Left: linear evolution. Right: fan diagrams summed by the nonlinear GLR equation (\ref{eq:glr1}).}
\label{fig:FanDiagram}
\end{figure}

The GLR equation sums a  subset of diagrams within  the double leading logarithmic approximation. They are called {\it fan diagrams},
 and are illustrated on the right hand graph in Fig.~\ref{fig:FanDiagram}.

The nonlinear Balitsky-Kovchegov  (BK) equation  is valid in the leading logarithmic $\ln 1/x$ approximation.
It has been derived independently by Kovchegov \cite{Kov} within the dipole formulation
of high energy scattering and by Balitsky \cite{Balitsky} from the operator product
expansion for high energy scattering. More precisely, Balitsky's equations form an infinite hierarchy
of coupled equations for correlators of Wilson lines, and only in the mean field approximation the first
equation decouples and is equivalent to the equation derived by Kovchegov.
An independent approach is that of Color Glass Condensate \cite{CGC} in which the evolution is governed by the 
JIMWLK functional equation \cite{JIMWLK} equivalent to Balitsky hierarchy.
In this lecture we will study the solution of the Balitsky-Kovchegov equation which is currently
the simplest tool to describe the parton saturation phenomenon.

\section{Multiple scattering in dipole picture}

\begin{figure}[htb]
\centerline{\epsfig{file=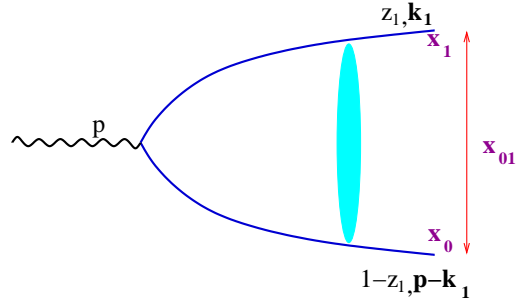,height=4cm}}
\caption{Heavy quark-antiquark dipole {\it onium}.}
\label{fig:Dipole0}
\end{figure}

In this section we will follow the derivation of the BFKL \cite{Mueller} and BK \cite{Kov} equations in the dipole picture.
Consider a heavy quark-antiquark pair, {\it onium}, shown in Fig.~\ref{fig:Dipole0}, who's wave function in the momentum space is denoted by
$$
\psi^{(0)}_{\alpha \beta}(k_1,z_1) \; ,
$$
where $k_1$ is the transverse momentum of the quark and $z_1=\frac{k_{1+}}{p_+}$ is the fraction of light cone momentum carried by the quark. 

The dipole picture is  formulated by  going to the transverse coordinate space
$$
{\psi^{{ (0)}}_{\alpha \beta}({ \xbzero},{\xbone},z_1) \, = \, \int \frac{d^2 \kbone}{(2\pi)^2}
{e^{i \xzerone \cdot \kbone} } \psi^{{(0)}}_{\alpha \beta}(\kbone,z_1) } \; ,
$$
$$
{\Phi^{(0)} (\xbzero,\xbone,z_1) \; = \; \sum_{\alpha,\beta} |\psi^{ (0)}_{\alpha \beta}(\xbzero,\xbone,z_1)|^2  } \; ,
$$
where  $\xbzero,\xbone$ denote the positions of the quark and antiquark respectively, which form the  end points of the {\it dipole}.\\

\begin{figure}[htb]
\centerline{\epsfig{file=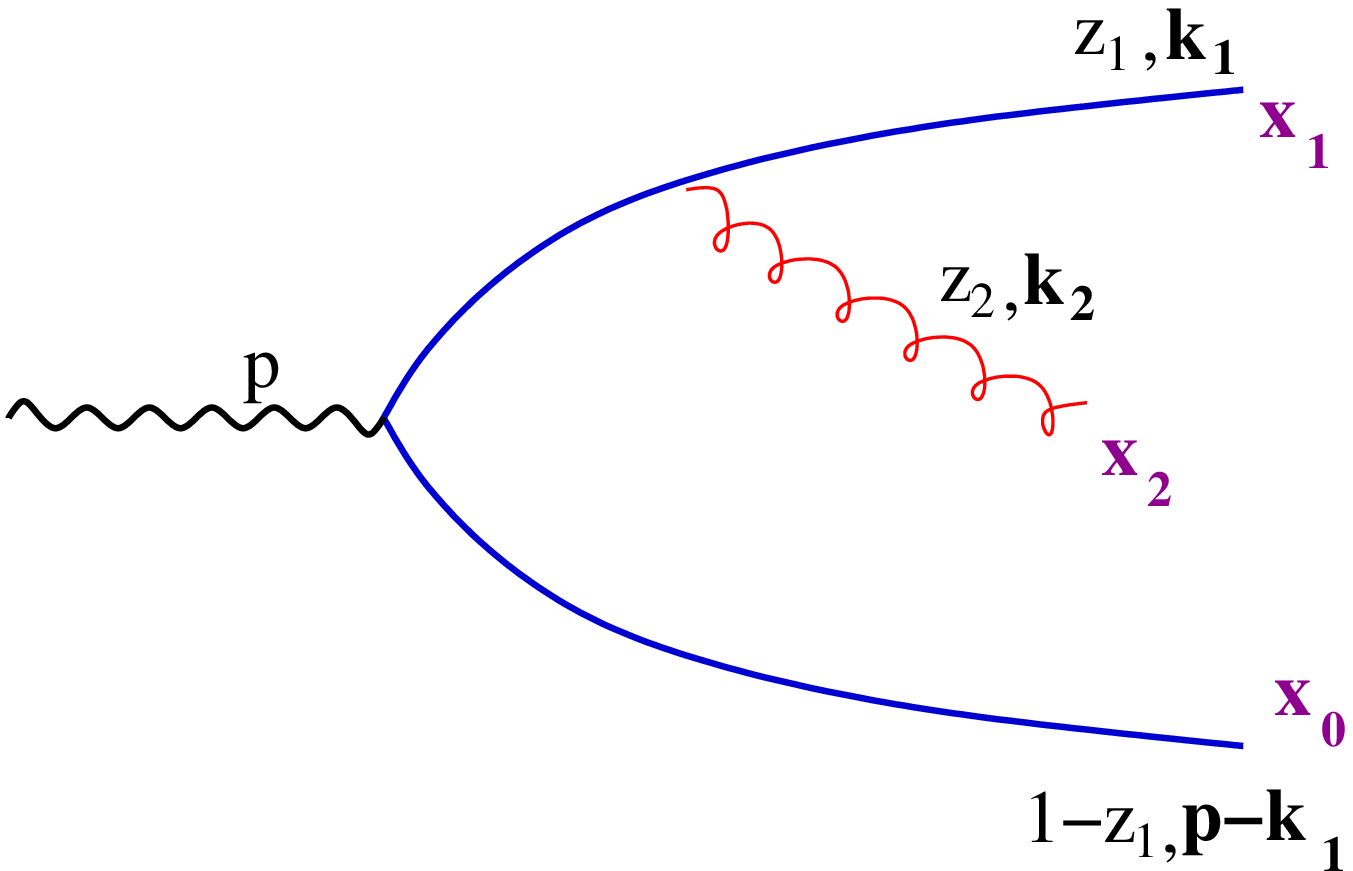,height=3.5cm}\hfill
\epsfig{file=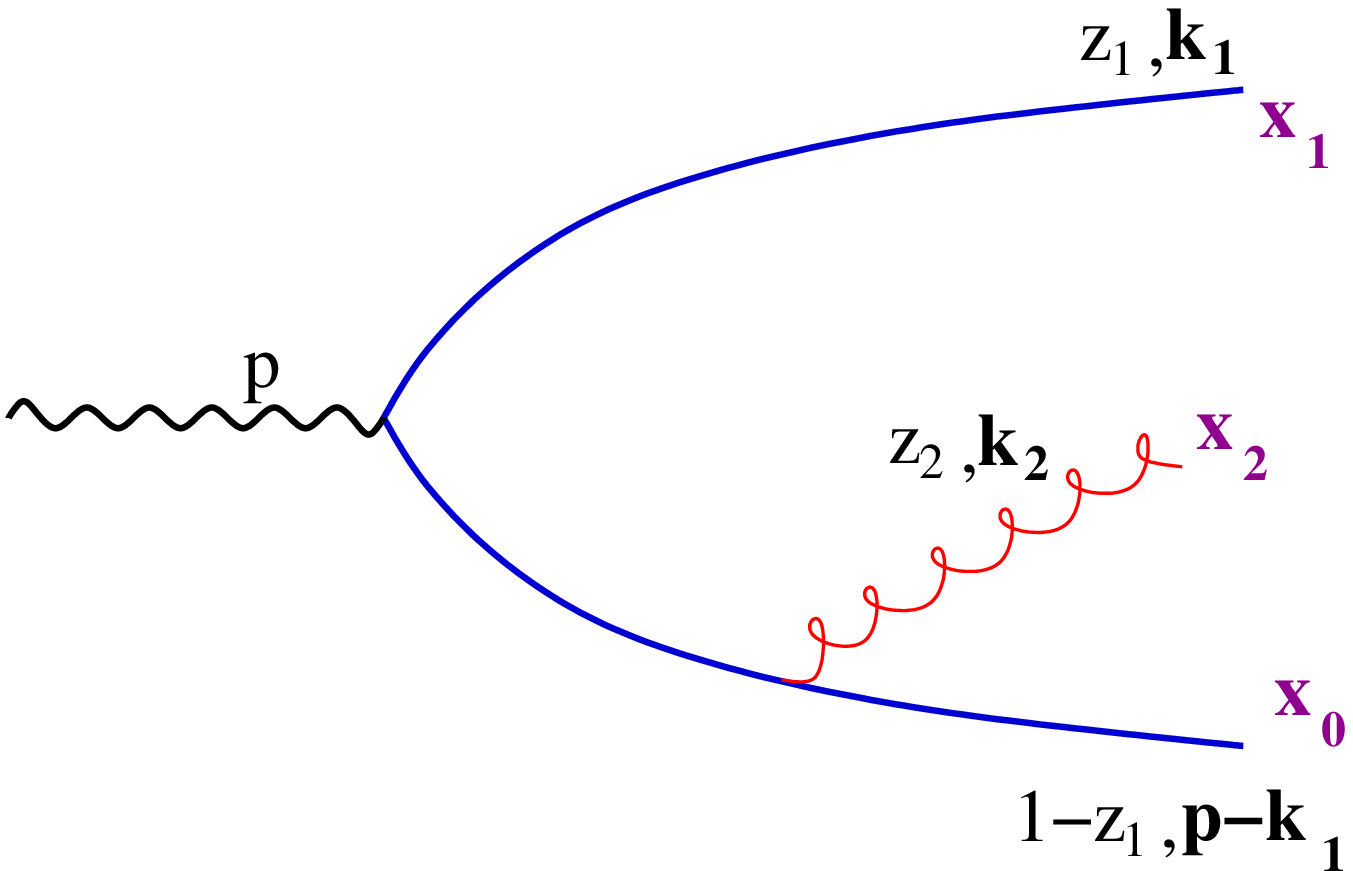,height=3.5cm}}
\caption{Onium with an additional single soft gluon.}
\label{fig:Dipole1}
\end{figure}

Then, as shown in Fig.~\ref{fig:Dipole1}, one adds one soft gluon with its longitudinal momentum  much smaller than that
of the original quark (anti-quark) $z_2/z_1 \ll 1$ .

\begin{figure}[htb]
\centerline{\epsfig{file=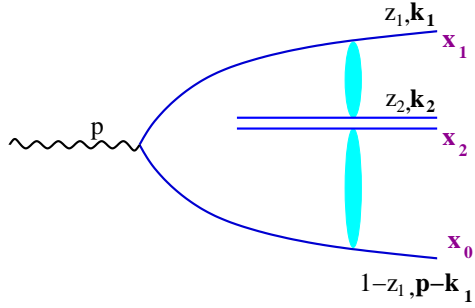,height=4cm}}
\caption{Onium wave function which consists of two dipoles.}
\label{fig:TwoDipoles}
\end{figure}

The relation between the  one-gluon wave function $\Phi^{(1)}$ and onium wave function  without any soft gluons $\Phi^{(0)}$
is 
$$
{\Phi^{{(1)}} (\xbzero,\xbone,z_1) \, = \, \frac{\alpha_s C_F}{\pi^2} \int_{z_0}^{z_1} \frac{dz_2}{z_2} \int {d^2 \xbtwo  \frac{\xzerone^2}{\xtwozer^2 \xonetwo^2}} \Phi^{{(0)}} (\xbzero,\xbone,z_1)} \; .
$$
In the limit of large number of colors, the gluon can be represented by a quark-antiquark pair, as in Fig.~\ref{fig:TwoDipoles}.
The  emission of one additional gluon is equivalent to the splitting of the original dipole $(0,1)$
into two dipoles $(0,2)$ and $(2,1)$ with probability of branching given by the measure
 $$ d^2 \xbtwo \frac{\xzerone^2}{\xtwozer^2 \xonetwo^2} \; . $$

\begin{figure}[htb]
\centerline{\epsfig{file=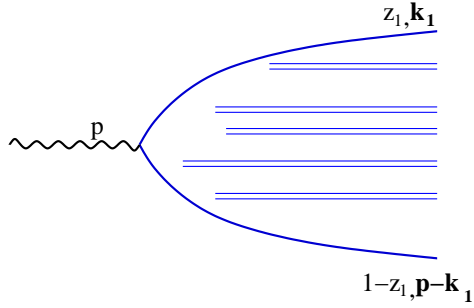,height=4cm}}
\caption{Onium wave function with arbitrary number of dipoles.}
\label{fig:Dipolen}
\end{figure}

The process of emissions of subsequent  soft dipoles can be repeated in the analogous way
to obtain the wave function with an arbitrary number of gluons $\Phi^{(n)}$, see Fig.~\ref{fig:Dipolen}. To describe such a process, Mueller \cite{Mueller}
 introduced
a dipole generating functional 
$$
Z({\bf b}_{01},{\bf x}_{01},z_1,u) \; ,
$$
which satisfies the normalization condition
$$
Z({\bf b}_{01},{\bf x}_{01},z_1,u=1)\;=\; 1 \; .
$$
The wave functions for arbitrary number of gluons can be obtained
by performing functional differentiation of $Z$,
$$
{
\hspace*{-0.2cm} \Phi^{(n)}({\bf x}_0,{\bf x}_1,{\bf x}_2,\dots,{\bf x}_{n+1})  =   \Phi^{(0)}\frac{\delta}{\delta u({\bf x}_2)}\frac{\delta}{\delta u({\bf x}_3)}\dots\frac{\delta}{\delta u({\bf x}_{n+1})} { Z({\bf x}_0,{\bf x}_1,z_1,{ u})}|_{ u=0}} \; .
$$
Here $\Phi^{(n)}$  gives probability of finding $n$ daughter dipoles 
that  originate from parent quark-antiquark dipole $(0,1)$.
The daughter dipoles  are produced  in positions $x_k$ with $k=2,\dots,n$.
In the following, we will use another  notation with
$$
{\bf x}_{01}\equiv \xbzero-\xbone \; ,
$$
representing the transverse size of the dipole
and
$$
{\bf b}_{01}\equiv \frac{\xbzero+\xbone}{2} \; ,
$$
the impact parameter (position) of this dipole.

\begin{figure}[htb]
\centerline{\epsfig{file=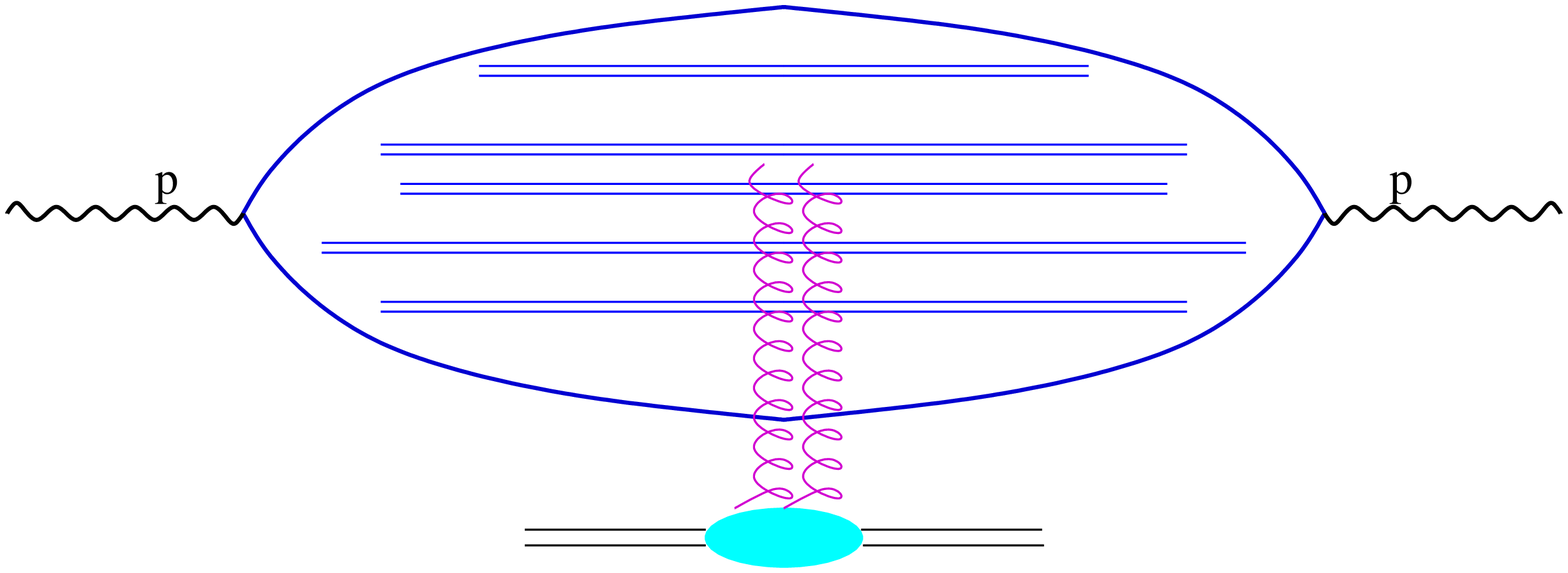,width=5.9cm}\hfill\epsfig{file=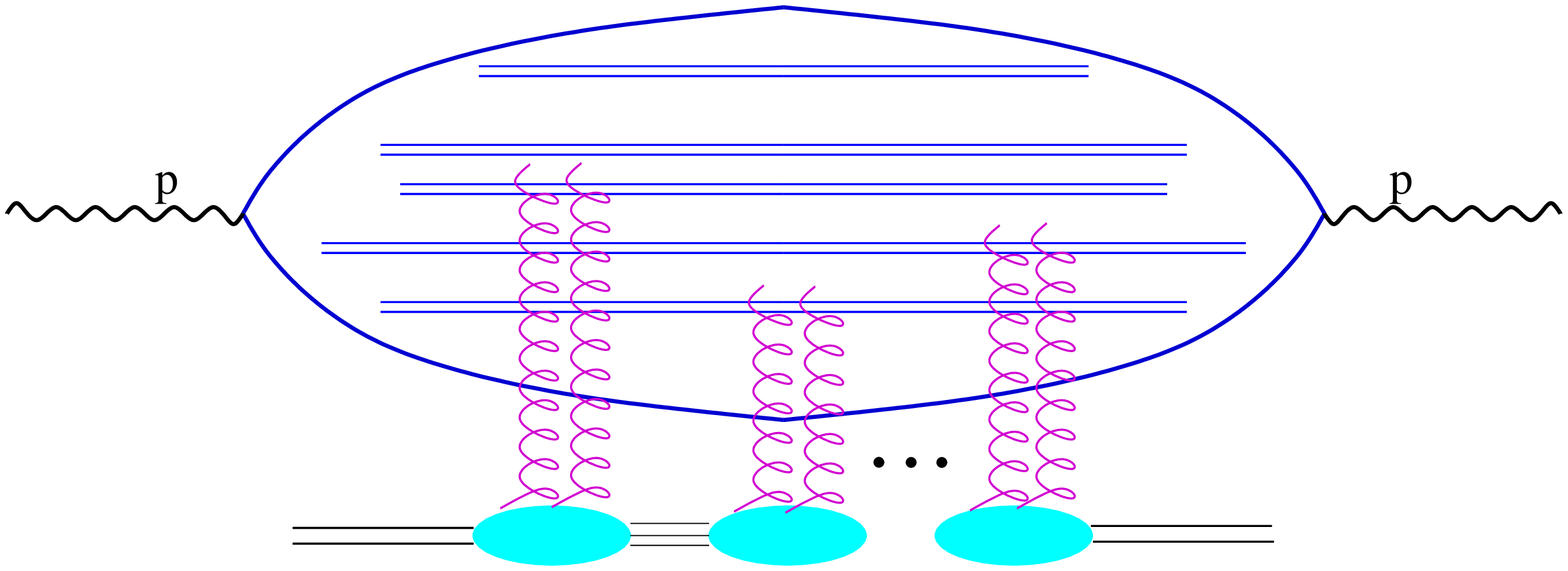,width=5.9cm}}
\caption{Left: single dipole scattering which leads to linear BFKL evolution equation (\ref{eq:BFKLDipole}). Right: multiple dipole scattering
which results in nonlinear Balitsky-Kovchegov evolution equation (\ref{eq:BKDipole}).}
\label{fig:DipScat}
\end{figure}

By investigating the relation between wave functions with $n$ and $n+1$ dipoles Mueller derived \cite{Mueller}
the following differential equation for the generating functional
\begin{multline}
\frac{dZ(\bzerone,\xzerone,y,u)}{dy}  =  \\
\int \frac{d^2 {\bf x}_2 \xzerone^2}{\xtwozer^2 \xonetwo^2} \,
\left[ Z(\bzerone+\frac{\xonetwo}{2},\xtwozer,y,u)\, Z(\bzerone-\frac{\xtwozer}{2},\xonetwo,y,u) \, - \, Z(\bzerone,\xzerone,y,u) \right] \; ,
\label{eq:GenFun}
\end{multline}
where the evolution variable is the rapidity $y=\ln 1/z_+$.
Using   equation (\ref{eq:GenFun}), one can obtain the evolution equation for the scattering amplitude
of dipole on the target.
First, one constructs the dipole number density  
$$
{ {n_1(x_{01},{\bf x},{\bf b}-{\bf b}_0,Y)} = \frac{\delta}{\delta u({\bf b},{\bf x})} Z(\bzerone,\xzerone,Y,u)|_{u=1}} \; ,
$$
and in general, the  density  for $k$ dipoles
\be
{n_k = \Pi_{i=1}^{k} \frac{\delta}{\delta u({\bf b}_i,{\bf x}_i)} \, Z|_{{u=1}}} \; .
\label{eq:nk}
\ee
The amplitude for scattering of single dipole on a target, see left graph in Fig.~\ref{fig:DipScat}, can be then obtained by convoluting
the dipole number density with the propagator of that dipole in the nucleus
\be
N_1(\xzerone,\bzerone,Y) = \int d[{\cal P}_1]  { n_1} \, \gamma_1 \; ,
\label{eq:amplitude1}
\ee
where ${ d[{\cal P}]_1 =  \frac{d^2{\bf x}_1}{2\pi x_i^2} d^2 {\bf b}_1} $ is the phase space measure, and  ${ \gamma \equiv\gamma({\bf x},{\bf b}) }$ is the propagator of a single dipole in the nucleus.
By differentiating the equation for the generating functional and using the relation (\ref{eq:amplitude1})
one can obtain the linear evolution equation for the dipole-target amplitude

\begin{multline}
\hspace*{-0.6cm} \frac{dN_1(\bzerone,\xzerone,Y)}{dY}  = 
 \bar{\alpha}_s \!  \int   \frac{d^2 {\bf x}_2 \, \xzerone^2}{\xtwozer^2 \, \xonetwo^2} \,
 \bigg[ N_1(\bzerone+\frac{\xonetwo}{2},\xtwozer,Y)+ N_1(\bzerone-\frac{\xtwozer}{2},\xonetwo,Y) \,\\
 - \, N_1(\bzerone,\xzerone,Y) \bigg] \, .
\label{eq:BFKLDipole}
\end{multline} 
It has to be stressed that  only the contribution  from the {\it single scattering}
of one dipole on the target has been included in the derivation. The  equation (\ref{eq:BFKLDipole}) is a dipole version of the BFKL equation in the transverse coordinate space as derived in \cite{Mueller}.
One can also generalize this equation by taking into account a  multiple scattering of many dipoles on the target, see right graph in Fig.~\ref{fig:DipScat}. To this aim one takes the  number density of $k$ dipoles, Eq.~(\ref{eq:nk}),
and then convolutes it with $k$ propagators for respective dipoles. The following expression for the amplitude is then 
\be
 N(\xzerone,\bzerone,Y) \, = \, \sum_{k=1}^{\infty} \int d[{\cal P}_k] \, {n_k} \, \gamma_1\dots\gamma_k  \; ,
\ee
where the measure is defined as
$$
[{\cal P}]_k = \Pi_{i=1}^{i=k} \frac{d^2{\bf x}_i}{2\pi x_i^2} d^2 {\bf b}_i \; .
$$
By differentiation of the equation for the generating functional $Z$ one can obtain the evolution equation
for the amplitude which takes into account also multiple scatterings \cite{Kov}
\begin{multline}
 \frac{dN(\bzerone,\xzerone,Y)}{dY}  = 
  \asb \int   \frac{d^2 {\bf x}_2 \, \xzerone^2}{\xtwozer^2 \, \xonetwo^2} \,
 \bigg[ N(\bzerone+\frac{\xonetwo}{2},\xtwozer,Y)+ N(\bzerone-\frac{\xtwozer}{2},\xonetwo,Y) \,\\
 - \, N(\bzerone,\xzerone,Y)  \, - \,{ N(\bzerone+\frac{\xonetwo}{2},\xtwozer,Y)\,N(\bzerone-\frac{\xtwozer}{2},\xonetwo,Y)} \bigg] \; .
\label{eq:BKDipole}
\end{multline} 
The characteristic feature of this equation is its nonlinearity. Thus, in the dipole approach, the multiple scattering of many dipoles in the onium leads to nonlinear evolution equation for the amplitude. This has to be contrasted with the single scattering  of one dipole which leads to the linear BFKL-type  equation.
One has to stress that this multiple scattering is a  completely incoherent process: dipoles scatter independently of each other and  there are no correlations. This is quite an important simplification which results in a  relatively simple and  closed evolution equation. These correlations are now a subject of intensive  research.
\section{Balitsky-Kovchegov equation at high energies}
\begin{figure}
\centerline{\epsfig{file=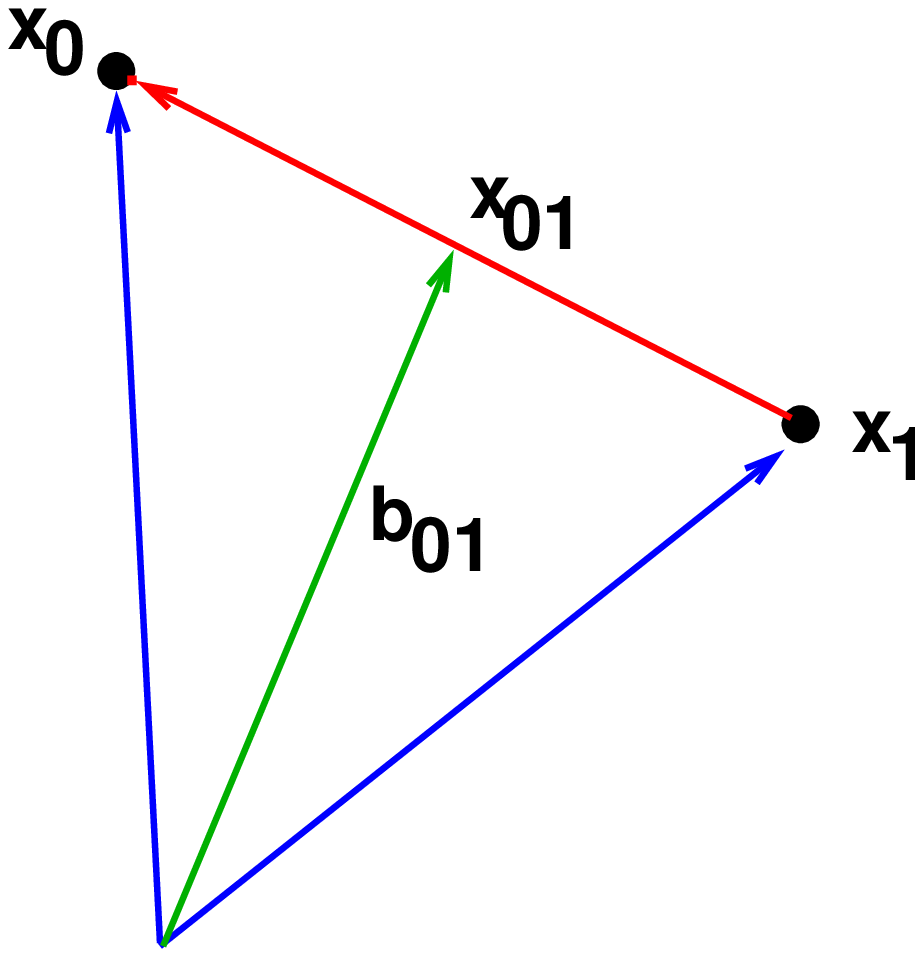,width=4cm}}
\caption{Schematic representation of the dipole position in impact parameter space.
$(\xbzero,\xbone)$ denote end points of the dipole.}
\label{fig:DipPos}
\end{figure}

In the remaining sections of the paper will be devoted to the discussion of  the solutions to the  BK equation
\begin{multline}
 \frac{dN(\bzerone,\xzerone,Y)}{dY}  = 
  \asb \int   \frac{d^2 {\bf x}_2 \, \xzerone^2}{\xtwozer^2 \, \xonetwo^2} \,
 \bigg[ N(\bzerone+\frac{\xonetwo}{2},\xtwozer,Y)+ N(\bzerone-\frac{\xtwozer}{2},\xonetwo,Y) \,\\
 - \, N(\bzerone,\xzerone,Y)  \, - \,{ N(\bzerone+\frac{\xonetwo}{2},\xtwozer,Y)\,N(\bzerone-\frac{\xtwozer}{2},\xonetwo,Y)} \bigg]  \; .
\label{eq:kov}
\end{multline} 
Let us highlit the salient features of this equation: \\
i) BK equation  is an evolution equation in rapidity $Y=\ln 1/x$. \\
ii) It requires the initial conditions $N^{(0)}(\bzerone,\xzerone,Y=0)$
which depend on the target  of the specific  process to be specified. \\
iii)  The BK  equation is valid in the leading logarithmic approximation in which the powers in $(\alpha_s \ln 1/x)^n$
are counted. \\
iv) In this approximation the strong coupling $\alpha_s$ is fixed. \\
v)  In (\ref{eq:kov}) $\bzerone$ represents the impact parameter whereas $\xzerone$ is  the size of the dipole, see Fig.~\ref{fig:DipPos}. The  problem  involves $(4+1)$ variables: four degrees of freedom per dipole and one evolution variable.\\

\subsection{Toy model in $(0+1)$ dimensions}
\begin{figure}[htb]
\centerline{\epsfig{file=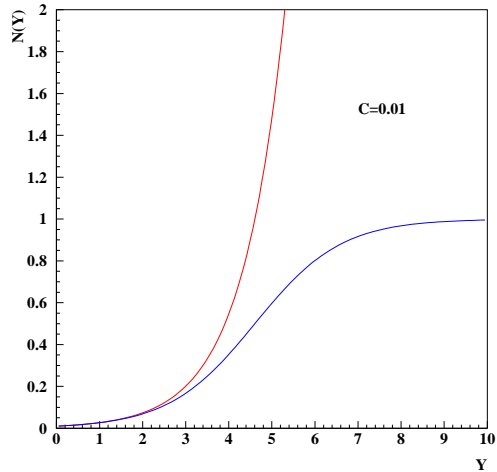,width=7.0cm}}
\caption{Illustration of the solution to the Verhulst equation (saturated line) and the linear equation (exponentially increasing).}
\label{fig:Verhulst}
\end{figure}

Searching  for the  solutions to Eq.~(\ref{eq:kov}) one has first to notice that  this equation has two fixed points
$$
\frac{d N(\bzerone,\xzerone,Y)}{dY} \; = \; 0  \; ,
$$
which are  at
$$
N=0 \hspace*{1cm} {\rm and} \hspace*{1cm}  N=1 \; .
$$
It is quite instructive to  first investigate  the toy model in $(0+1)$ dimensions, 
 when amplitude depends only  on the rapidity  $N\equiv N(Y)$ and the kernel is simply a constant.
Then  the equation reduces to 
$$
\frac{dN}{dY} \; = \; \omega (N-N^2)\, , \; \omega>0 \; .
$$
The above equation was first discussed by Verhulst in 1838 as a model for self-limiting population growth in biology.
The solution to this equation, called logistic curve, can be easily found
$$
N(Y) \; = \; \frac{e^{\omega Y}}{e^{\omega Y}+C^{-1}} \; , \hspace*{0.5cm} N^{(0)}(Y=0) = C \; ,
$$
and is illustrated in Fig.~\ref{fig:Verhulst}. 
Its  crucial property is that it saturates for very large values of $Y$
$$
\forall_{C\neq 0} \; N(Y) \stackrel{Y\rightarrow\infty}{\longrightarrow} 1 \; ,
$$
in contrast to the solution of the linear equation which grows exponentially.

The toy model teaches us that
the fixed point at $N=0$ is unstable  with respect to the linear part of the  evolution. In contradistinction, the value of $N=1$ turns out to be a fixed point. After sufficiently long  evolution in $Y$ the solution will reach this fixed point, independently of initial conditions (provided  that $N^{(0)}(Y=0) \neq 0$).

\section{Solution in $(1+1)$ dimensions}
Having briefly looked at the toy model let us proceed to the full equation
which depends on $(4+1)$  variables which makes it very  difficult to solve even numerically. The biggest complication is that the variables $\bzerone$ and  $\xzerone$ are entangled in the arguments of the functions $N$, see Eq.(~\ref{eq:kov}).
However,  the kernel depends only on the sizes $\xzerone,\xtwozer,\xonetwo$. By assuming that the solution $N$ is translationally  invariant
 $$N(\bzerone,\xzerone,Y)\rightarrow N(|\xzerone|,Y) \; ,$$
the problem is reducible to $(1+1)$ dimensions with no dependence on the impact parameter $\bzerone$. Physically, this corresponds to scattering of a dipole on an infinite and uniform nucleus.
In this approximation, the BK equation  in $(1+1)$ dimensions has been extensively studied numerically,
\cite{Braun,Armesto,Lublinsky,GBMS} and analytically \cite{LT,MunPesch,BFL}.

\begin{figure}
\centerline{\epsfig{file=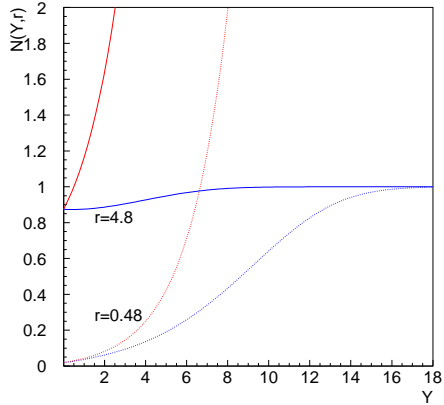,width=6.0cm}}
\caption{Rapidity dependence of the solution to $(1+1)$ dimensional Balitsky-Kovchegov equation (saturated lines) for two, fixed values of the dipole size $r$ as compared to the solution of the linear BFKL equation (exponentially increasing). }
\label{fig:Soll11Y}
\end{figure}

In Fig.~\ref{fig:Soll11Y} we illustrate the rapidity dependence of $N(Y,r=|\xzerone|)$ for two fixed  values of $r$, and compare it to the solution of the linear equation.
We notice that the solution to the BK equation has the same qualitative  features as the toy model. For any given $r$ the solution of the nonlinear equation tends to unity, whereas the linear solution  increases exponentially.

\begin{figure}
\centerline{\epsfig{file=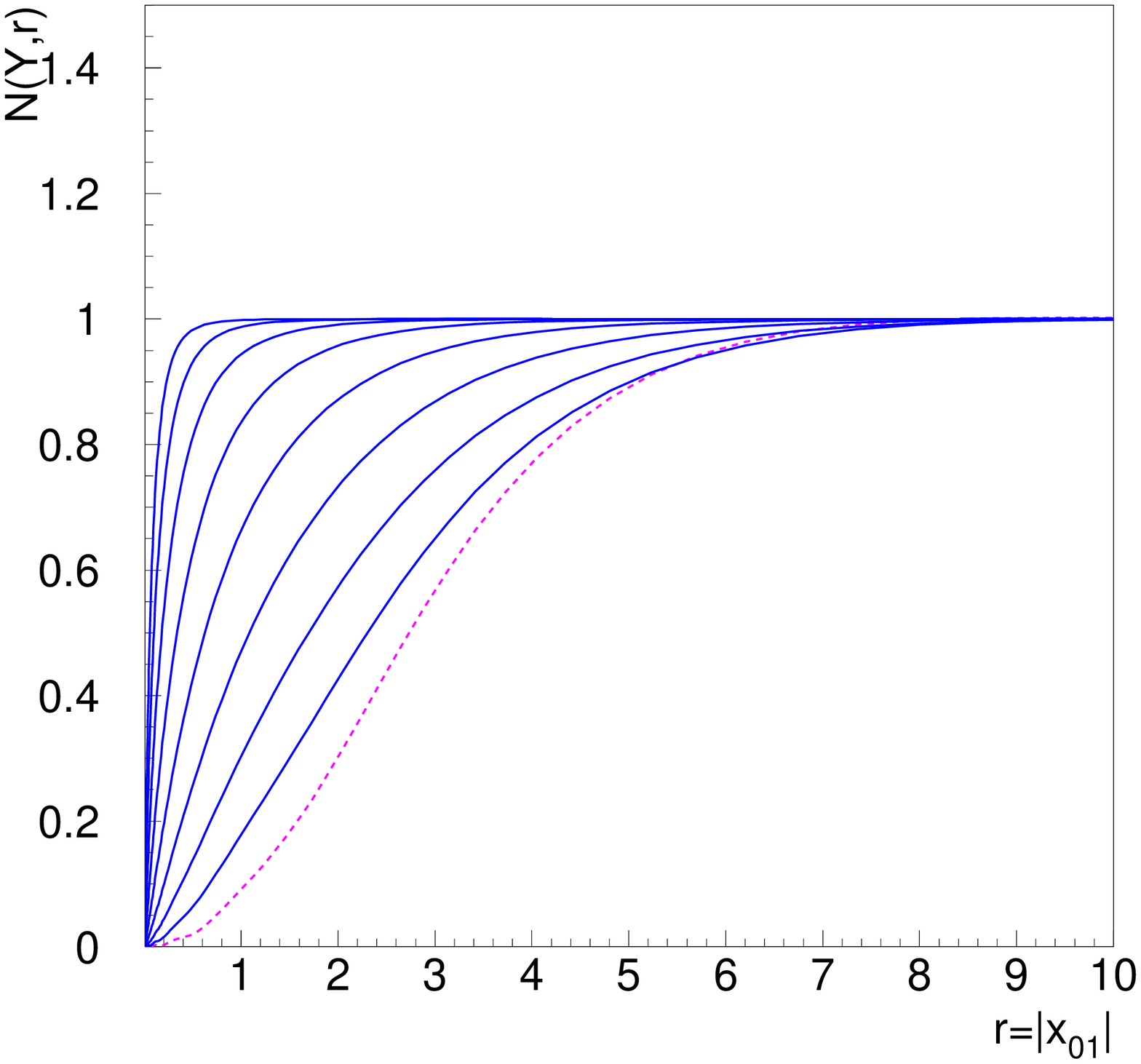,width=4.7cm}\hspace*{1cm}\epsfig{file=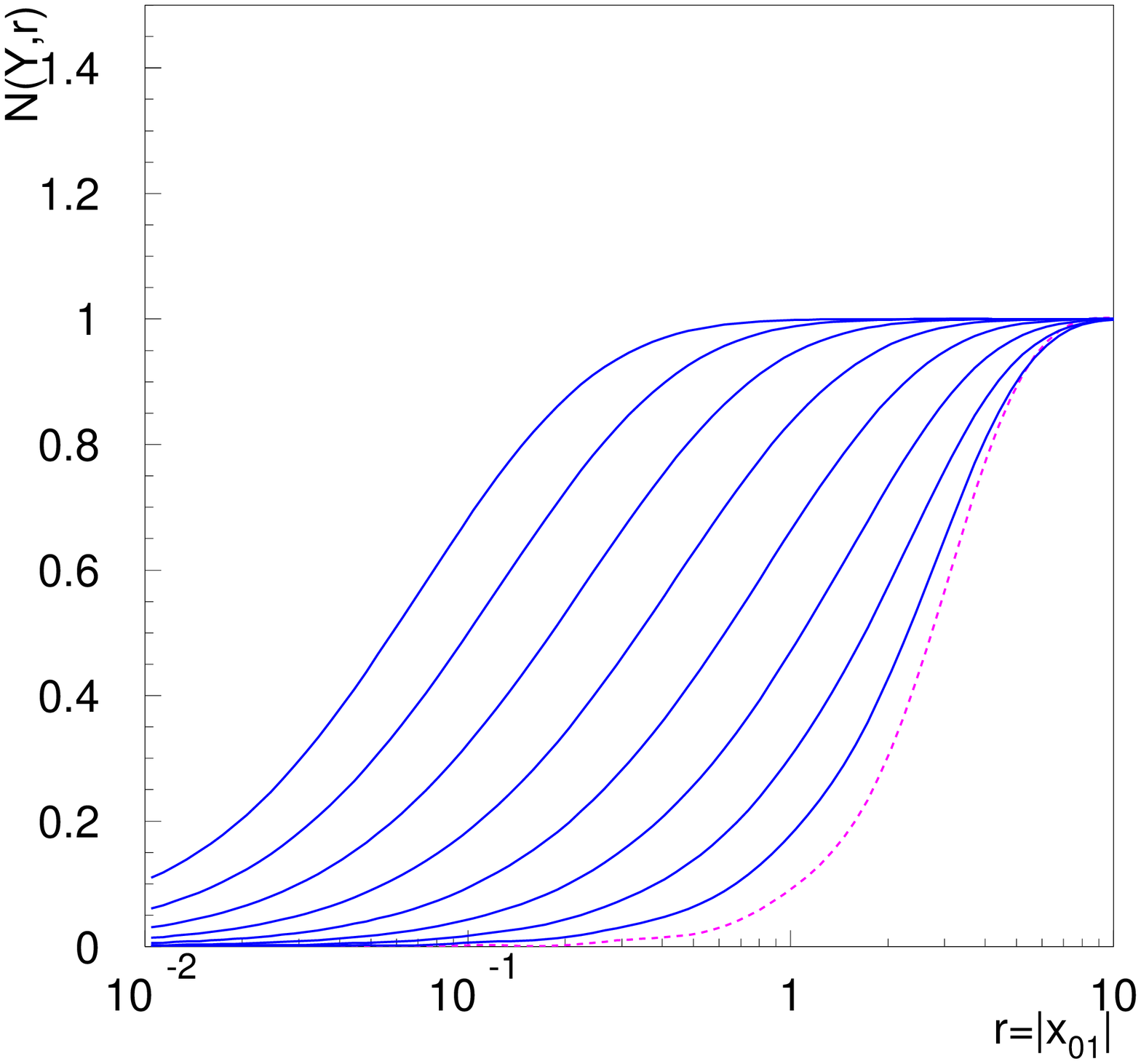,width=4.7cm}}
\centerline{\epsfig{file=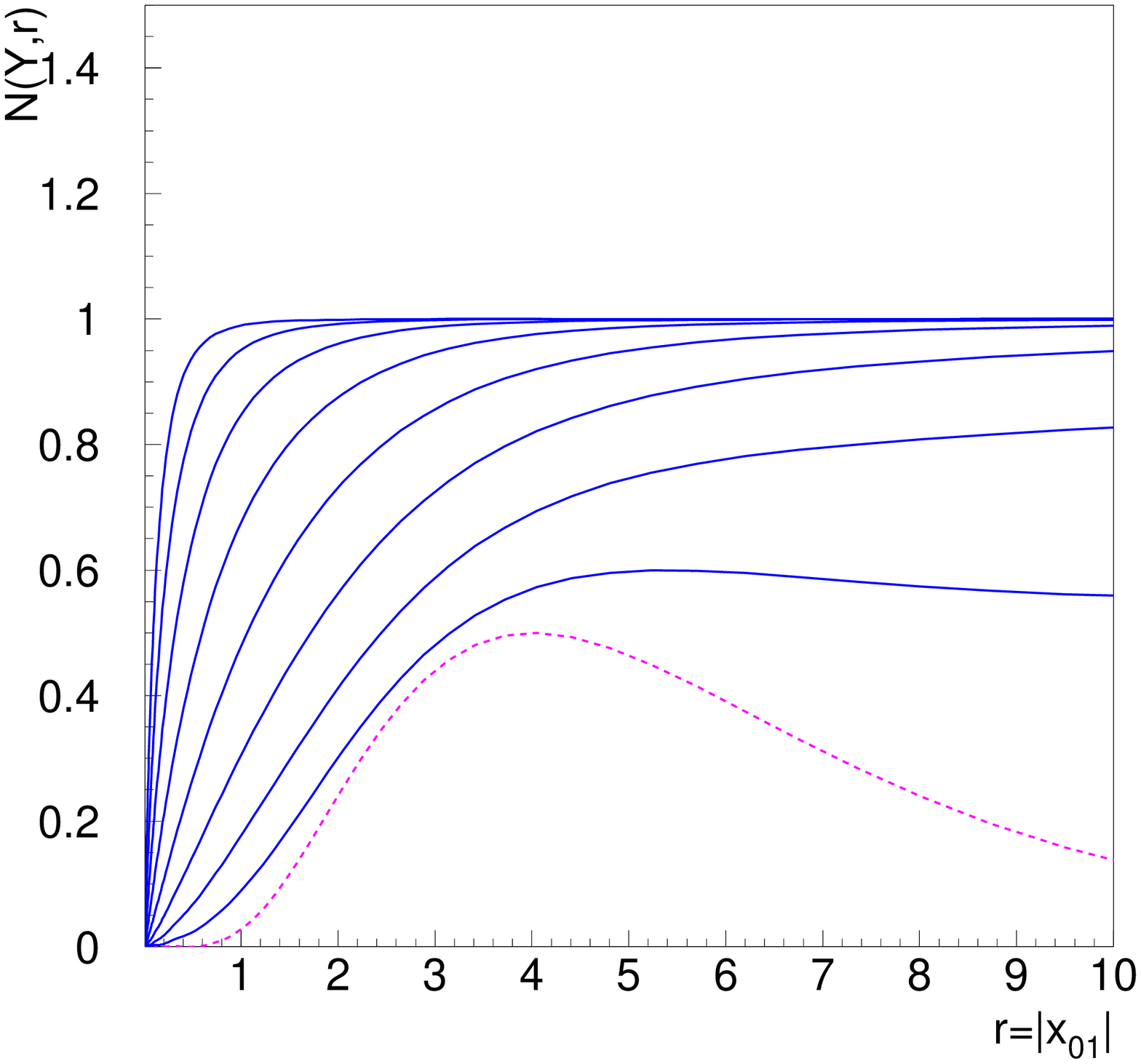,width=4.7cm}\hspace*{1cm}\epsfig{file=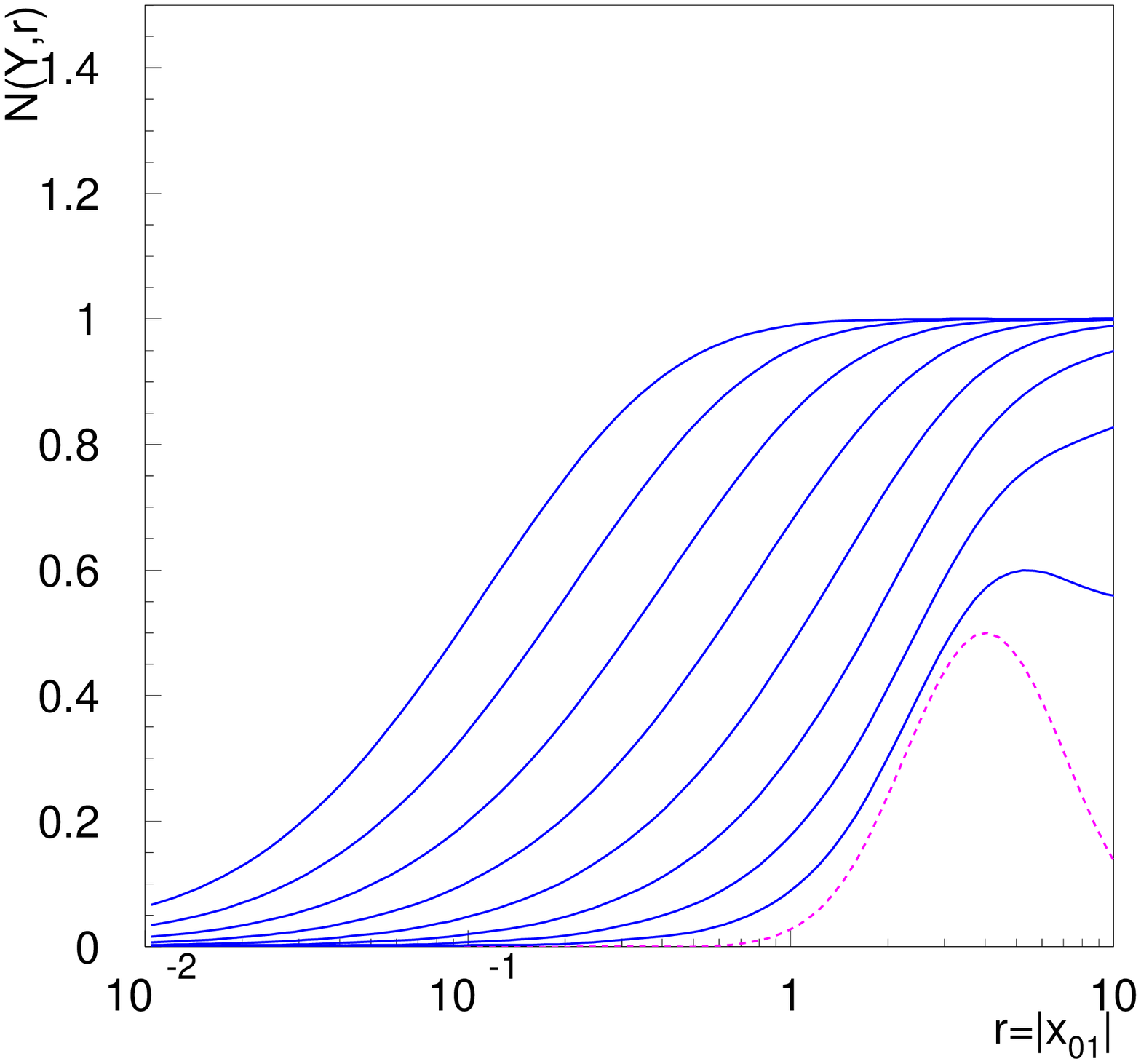,width=4.7cm}}
\centerline{\epsfig{file=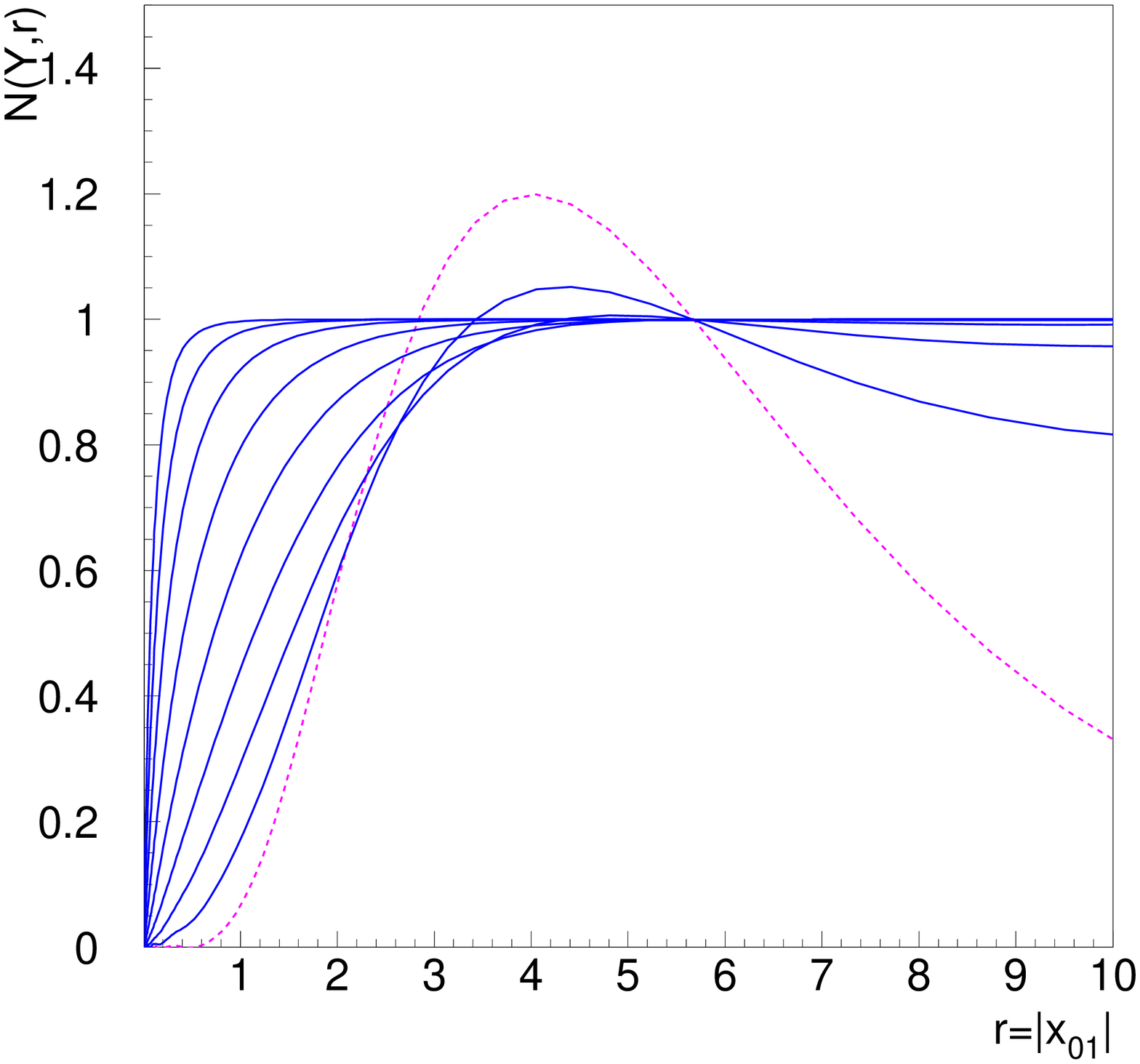,width=4.7cm}\hspace*{1cm}\epsfig{file=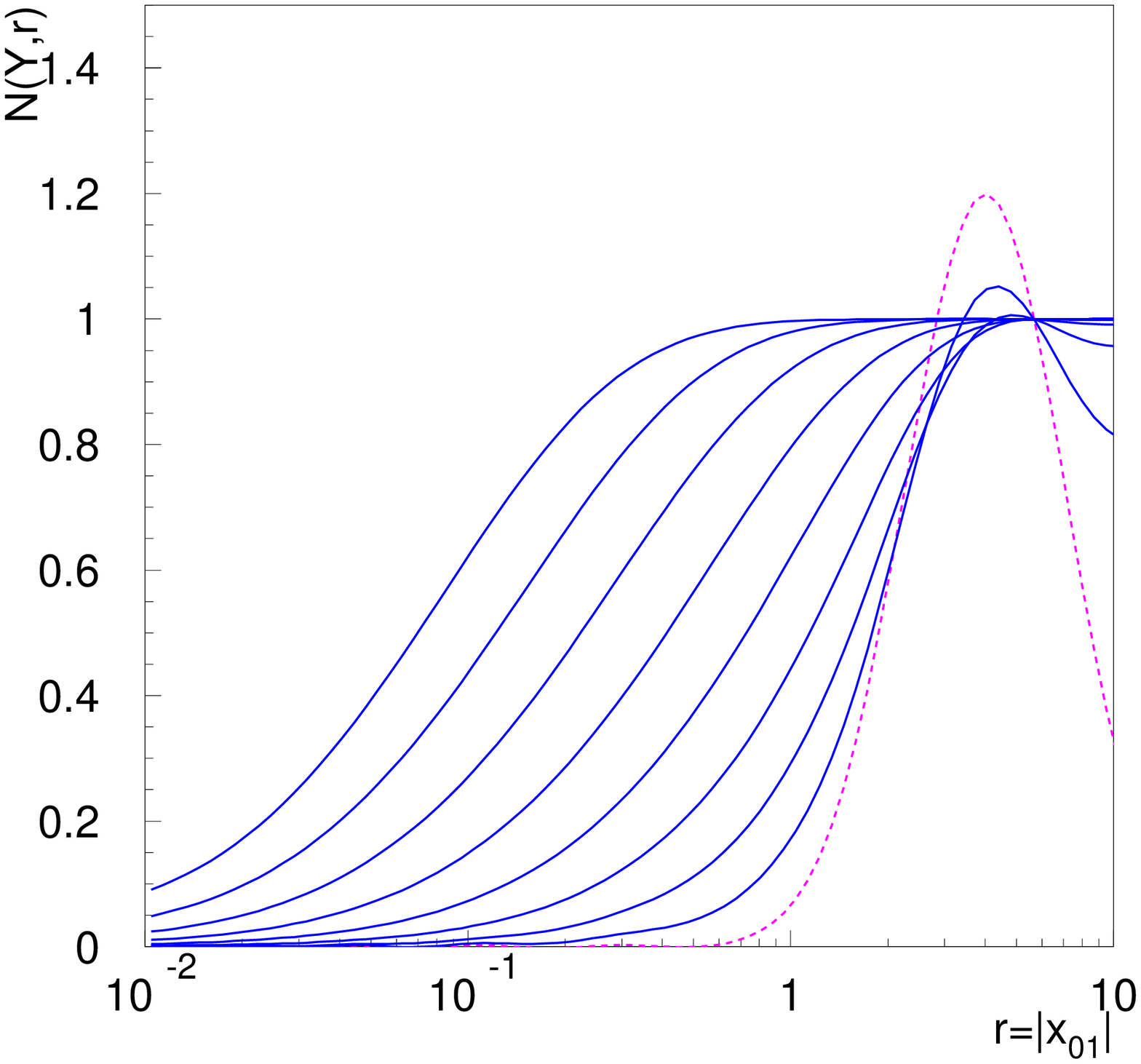,width=4.7cm}}
\caption{Dipole size dependence of the solution to the $(1+1)$ dimensional Balitsky-Kovchegov equation for different values of rapidity. From top to bottom: different initial conditions; left: linear scale; right: logarithmic scale. Dashed line denotes the initial distribution at $Y=0$. Solid lines from right to left are for increasing values of rapidity.}
\label{fig:InCon}
\end{figure}

Also  when the dipole size is larger, the system also saturates earlier,  for smaller values of $Y$.
To understand this feature better let us  look at the solution as a function of the dipole size $r$ for given, fixed values of $Y$.
We notice  that for smaller dipoles the amplitude $N =1 $  as the rapidity increases. In the following, see Fig.~\ref{fig:InCon},
we will also study  the dependence of $N$ on  different initial conditions.

In Fig.~\ref{fig:InCon} several initial conditions have been shown.
They require different normalisations  and exhibit  different type of behaviour for large values of $r$ ($N(r)\rightarrow 1$ or $N(r)\rightarrow 0$ as $r\rightarrow \infty$). In all cases the common feature of the solution is that
 $N=0$ is an unstable fixed point and $N=1$ is the stable one.

\subsection{Saturation scale}
The solutions shown in Fig.~\ref{fig:InCon} exhibit three different behaviours : in one  region where the amplitude is small
 the nonlinear corrections are  negligible, the transition region and the asymptotic region 
 where the amplitude $N\sim 1$.
 The boundary of the transition region is characterized  by a saturation scale $Q_s(Y)$
\bea
r < \frac{1}{Q_s(Y)} & \rightarrow & N\ll 1 \hspace*{0.3cm} {\rm to the left} \; , \nonumber \\
r > \frac{1}{Q_s(Y)} & \rightarrow & N\sim 1 \hspace*{0.3cm} {\rm to the right} \; .\nonumber
\eea

The saturation scale can be extracted from the solution to the BK equation.
The leading rapidity dependence is exponential

$$
Q_s(Y) \; = \; Q_0 \exp({\bar\alpha}_s \,\lambda\, Y)\, Y^{-\beta} , \hspace*{1cm} \lambda \simeq 2.4 \; ,
$$

with some subleading  corrections  \cite{MuelTrian,MunPesch}.

The qualitative properties of the BK equation are roughly similar
to the properties of the dipole cross section of the  Golec-Biernat and Wusthoff saturation model \cite{GBW} in which the following form of the dipole cross section has been postulated
\be
\sigma(Y,r)\equiv \int d^2{\bf b} \, N({\bf b},r,Y) = \sigma_0 \left[1-\exp\left(-\frac{r^2 Q_s^2(Y)}{4}\right) \right] \; .
\label{eq:gbw}
\ee
In (\ref{eq:gbw})  the  saturation scale: $Q_s^2(Y) = e^{0.28 (Y-Y_0)}$ is rapidity dependent.
The  normalisation  $\sigma_0$ has been adjusted to obtain the best description of the experimentally measured proton structure function  $F_2$.
In the regime where dipoles are smaller than the reciprocal of the characteristic saturation scale $r<1/Q_s(Y)$, the cross section is small and proportional to
$$
\sigma(r,Y)/\sigma_0 \simeq r^2 Q_s^2(Y) /4 \; ,
$$
in accordance with the requirement of color transparency. When we consider large dipoles,
$r>1/Q_s(Y)$,  the cross section saturates to $\sigma_0$ 
$$
\sigma(r,Y) / \sigma_0 \simeq 1 \; ,
$$
and becomes independent of both $r$ and $Y$.

\subsection{Geometrical scaling and travelling waves}

While investigating the solution as  a function of the dipole size $r$ one observes that
the solution reaches the universal shape independently of the particulars of the  initial condition. 
For different values of  $Y$ the solutions have similar shapes but are shifted 
shifted towards smaller values of the dipole size.
This property is known as {\it geometrical scaling} and it was first  searched for and found in the  data at HERA electron-proton collider \cite{SGBK,GSother}.
Mathematically, geometrical scaling means that the solution to the BK equation depends
only on a single  combined variable 
$$
r Q_s(Y) \; ,
$$
instead of $r$ and $Y$ separately, i.e.
$$
N(r,Y) \, \equiv \, N(r Q_s(Y)) \; .
$$
In terms of logarithms of variables, using the rapidity dependence of the saturation scale $Q_s(Y) \simeq Q_0 \exp(\asb \lambda_s Y)$ one finds that 
$$
\ln r+ \ln Q_s(Y) = \ln r + \asb \lambda_s Y \; .
$$
If we interpret $\ln r$ as a spatial coordinate and $Y$ as the time variable, then the property of geometrical scaling implies that the solution is a  wave  front that propagates with a constant velocity $\asb \lambda_s$, see  \cite{MunPesch}. It has been also described as a  soliton wave \cite{Braun}.
The scaling property is also present in the Golec-Biernat and Wusthoff saturation model (\ref{eq:gbw}).

\begin{figure}
\epsfig{file=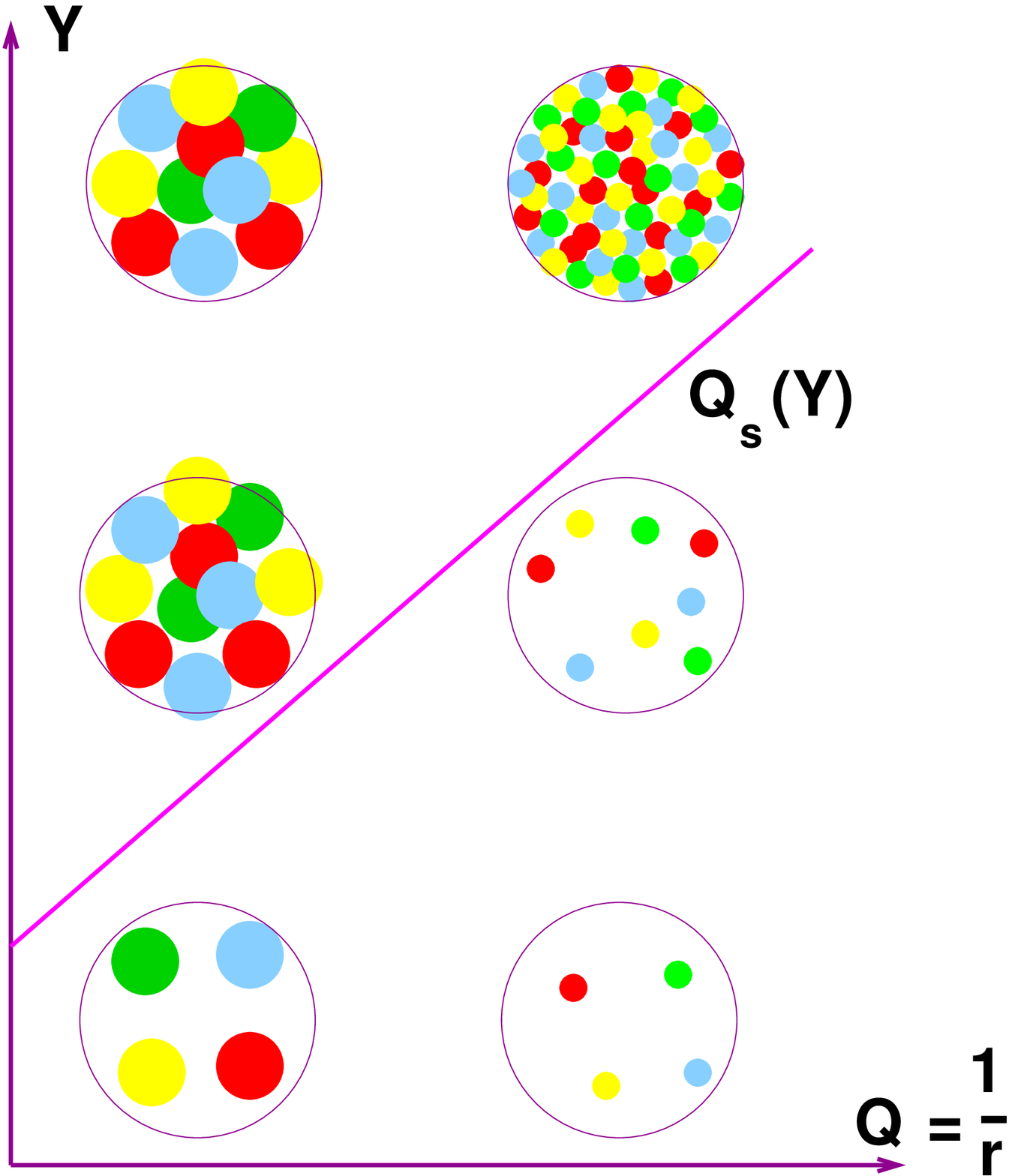,width=5cm}\hfill
\epsfig{file=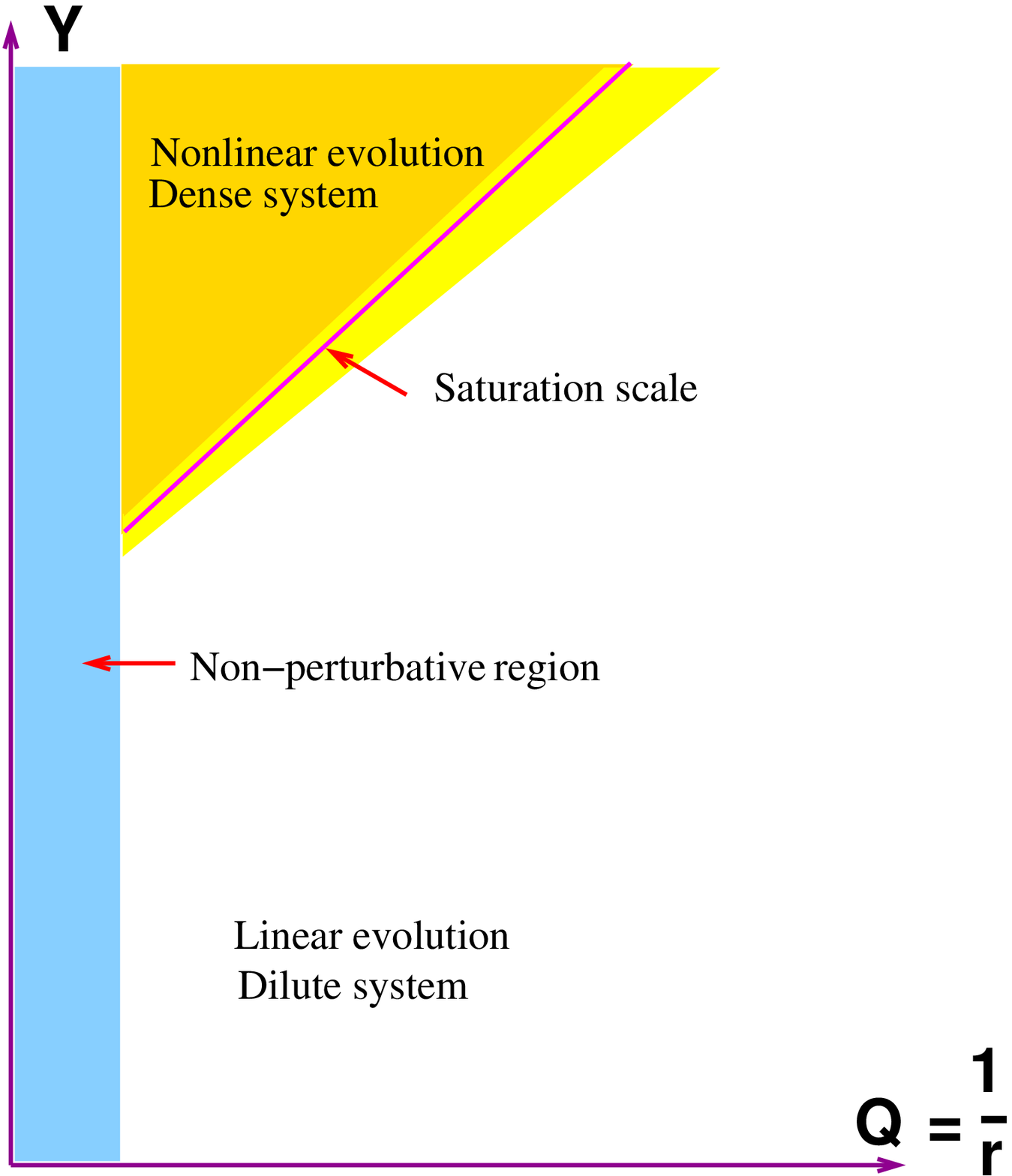,width=5cm}
\caption{Schematic representation of dilute and saturated regions regions in the kinematic space $(Q,Y)$.}
\label{fig:SatPic}
\end{figure}

The transition between the dilute and saturated regimes can be illustrated as in Fig.~\ref{fig:SatPic}. 
The dense and dilute regions are  divided by the critical line
 identified as  the saturation scale. 
The higher the rapidity the denser the  system becomes and finally partons begin to reinteract. The  saturation occurs earlier also, for  the larger  size of the partons.

\section{Diffusion properties of the BK equation}
Insofar we have looked at the solutions   in the coordinate space 
in accordance with the original formulation of the BK theory.
 By Fourier transform to the  momentum space 
$$
\phi({k},Y) := \int_0^{\infty} \frac{dr}{r} J_0({k\, r}) \, N({r},Y) \; ,
$$
in $(1+1)$ dimensions one can obtain
a more  compact form of this equation
\be
\frac{d \phi( k,Y)}{dY} \; = \; \asb \int \frac{dk'}{k'} {\cal K}(k,k') \, \phi(k',Y) \, - \, \asb \phi^2(k,Y) \; .
\label{eq:BKmom}
\ee
In Eq.~(\ref{eq:BKmom}) the integral operator ${\cal K}(k,k')$ is the usual BFKL kernel in momentum space.

\begin{figure}
\epsfig{file=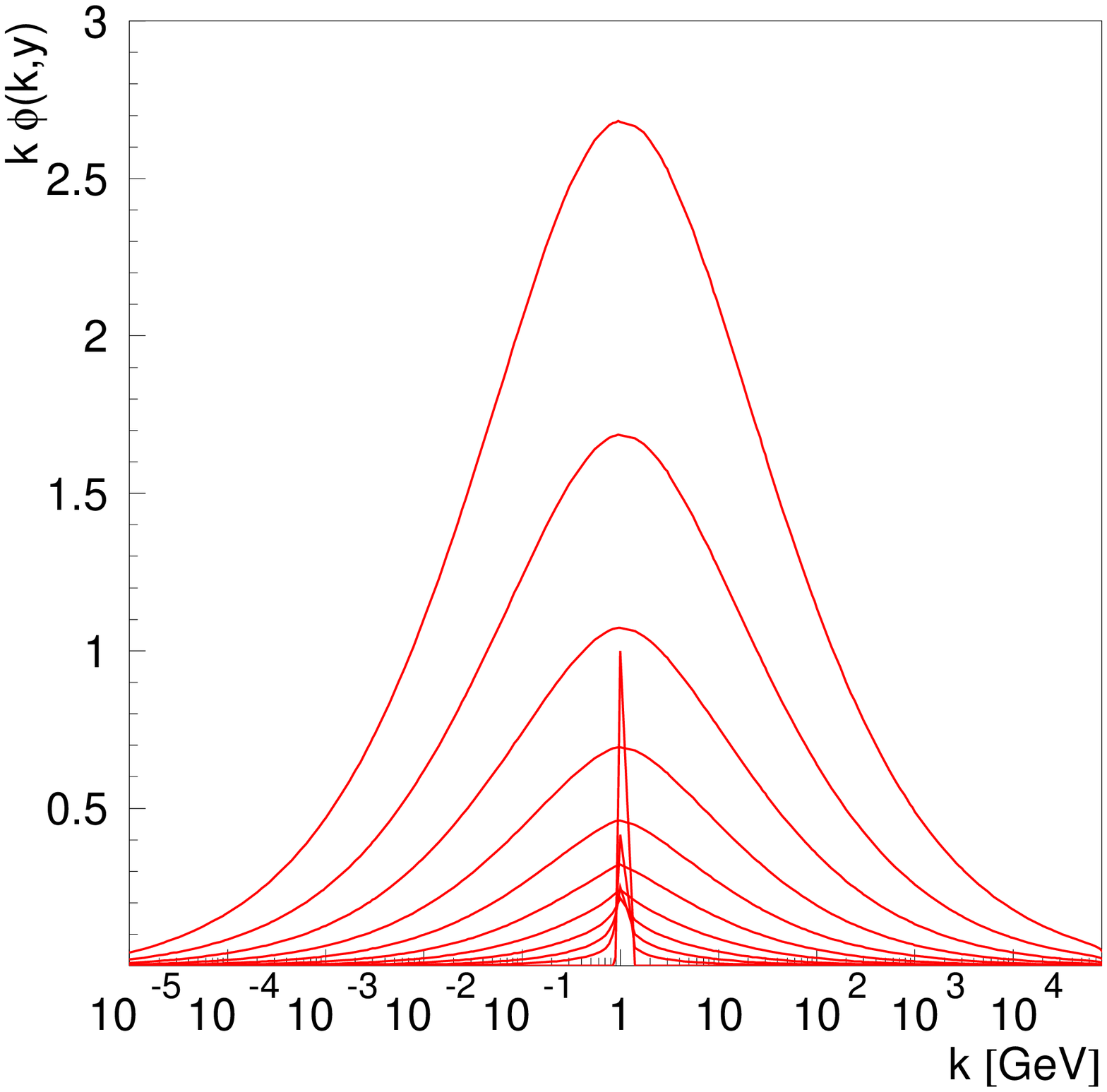,width=5.5cm}\hfill\epsfig{file=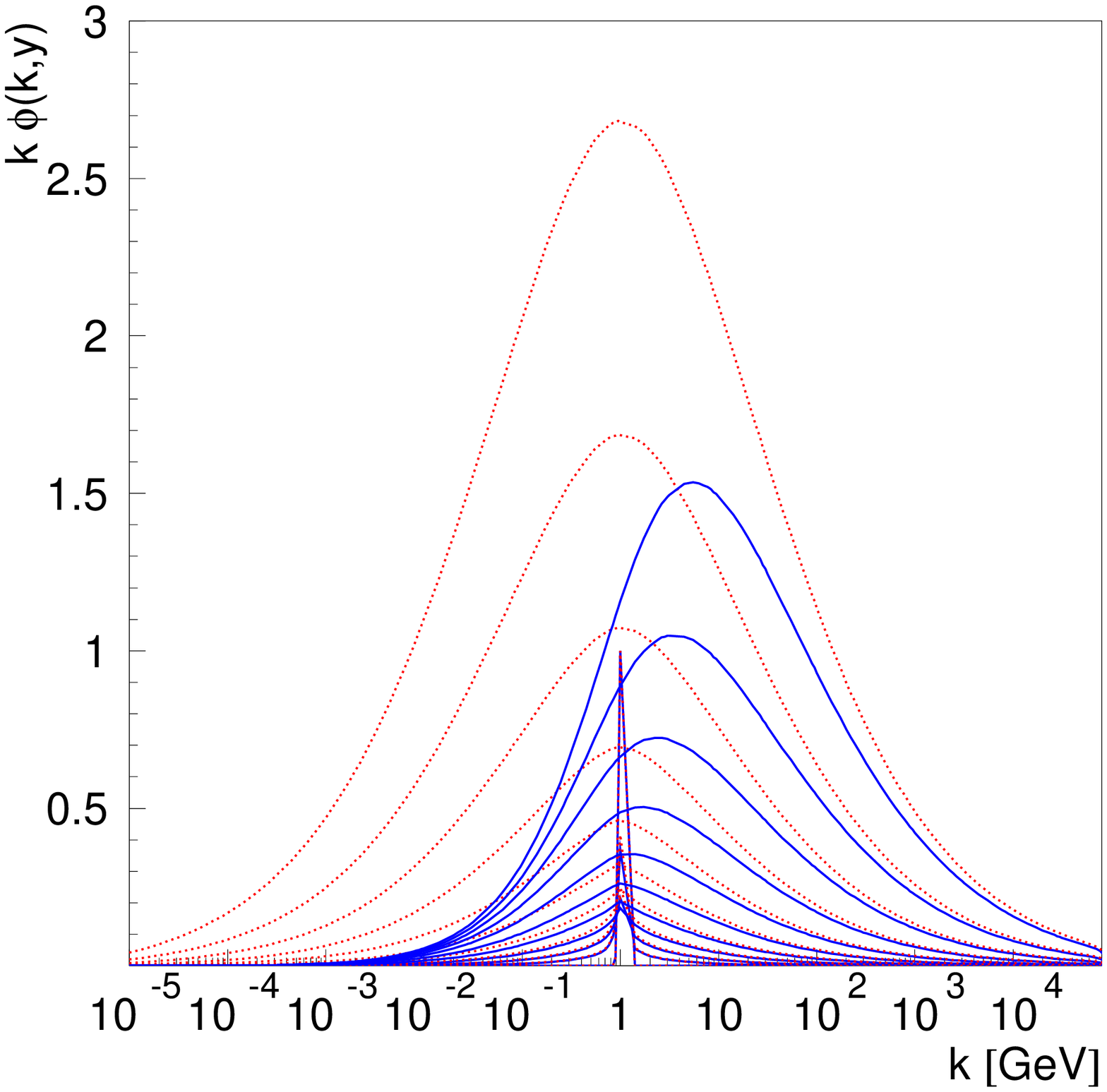,width=5.5cm}
\caption{Left: solution to the BFKL equation in the momentum space as a function of momentum $k$ for various fixed values of rapidity $Y=1,\dots,10$; right: the same but both BFKL (dashed) and BK (solid) solutions are showed.}
\label{fig:DiffBFKL}
\end{figure}

The solution to the linear part  is  well known.
In the saddle point approximation it reads

\be
k\phi(k,Y) \; = \; \frac{1}{\sqrt{\pi \asb \chi^{''}(0) Y}} \exp(\asb \chi(0) Y)\, \exp\left({-\frac{\ln^2 (k^2/k_0^2)}{2\asb \chi^{''}(0) Y}}\right) \; .
\label{eq:SaddleBFKL}
\ee
The first exponential is responsible for the fast increase of the gluon density  with rapidity. The value of $\chi(0)=4 \ln 2$ is the famous BFKL intercept. The second exponential
causes the diffusion of  the momenta into the ultraviolet and infrared regions.
It is well known fact that the BFKL equation exhibits  strong diffusion, which
can be interpreted as a random walk in the $\ln k$  space of transverse momenta.
The rapidity (energy) plays here the  role of the time variable. 
 The left  plot of Fig.~\ref{fig:DiffBFKL}  illustrates the numerical solution to the BFKL as a function of transverse momentum for fixed values of $Y$.

\begin{figure}
\epsfig{file=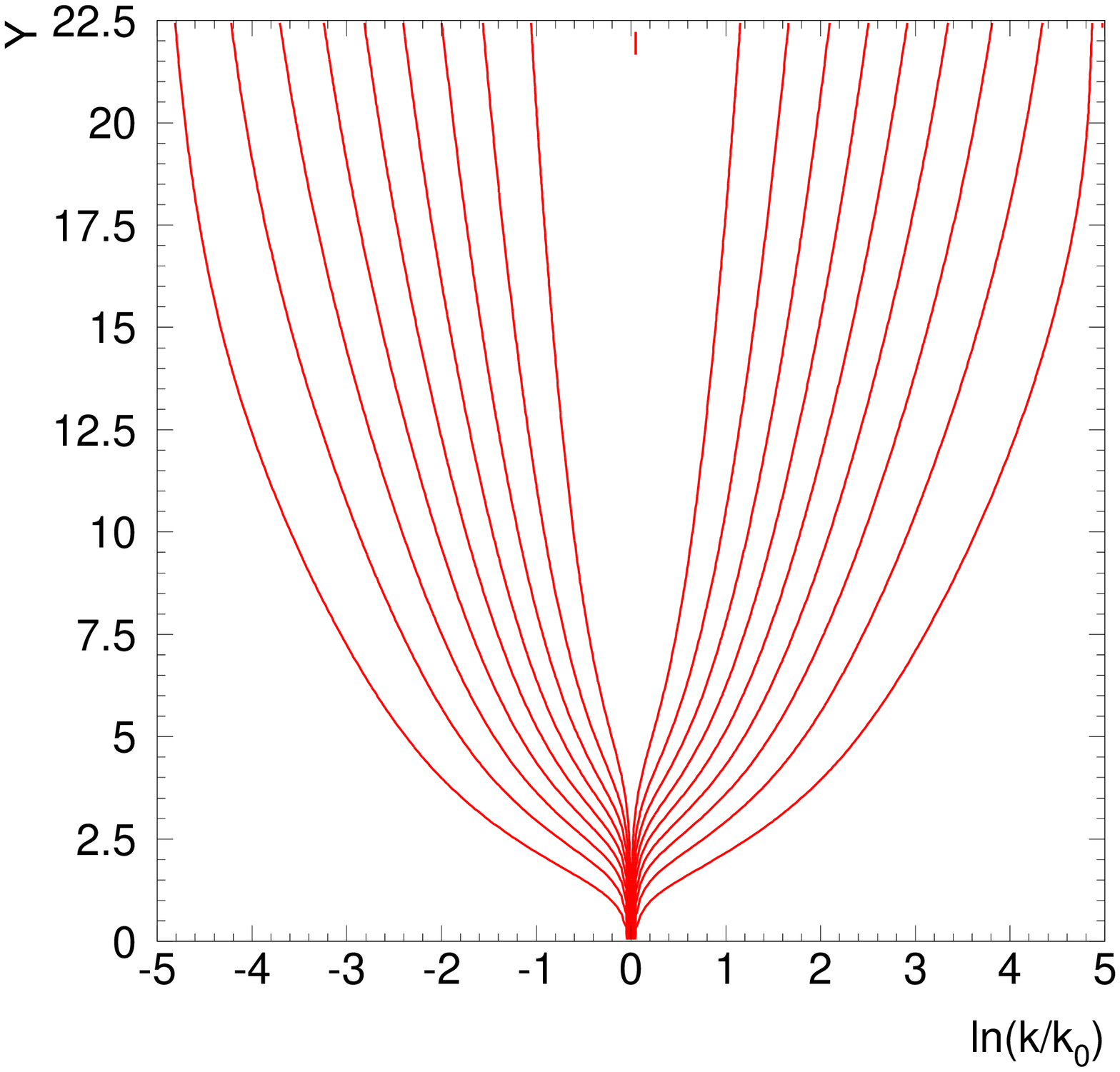,width=6.cm}\hfill\epsfig{file=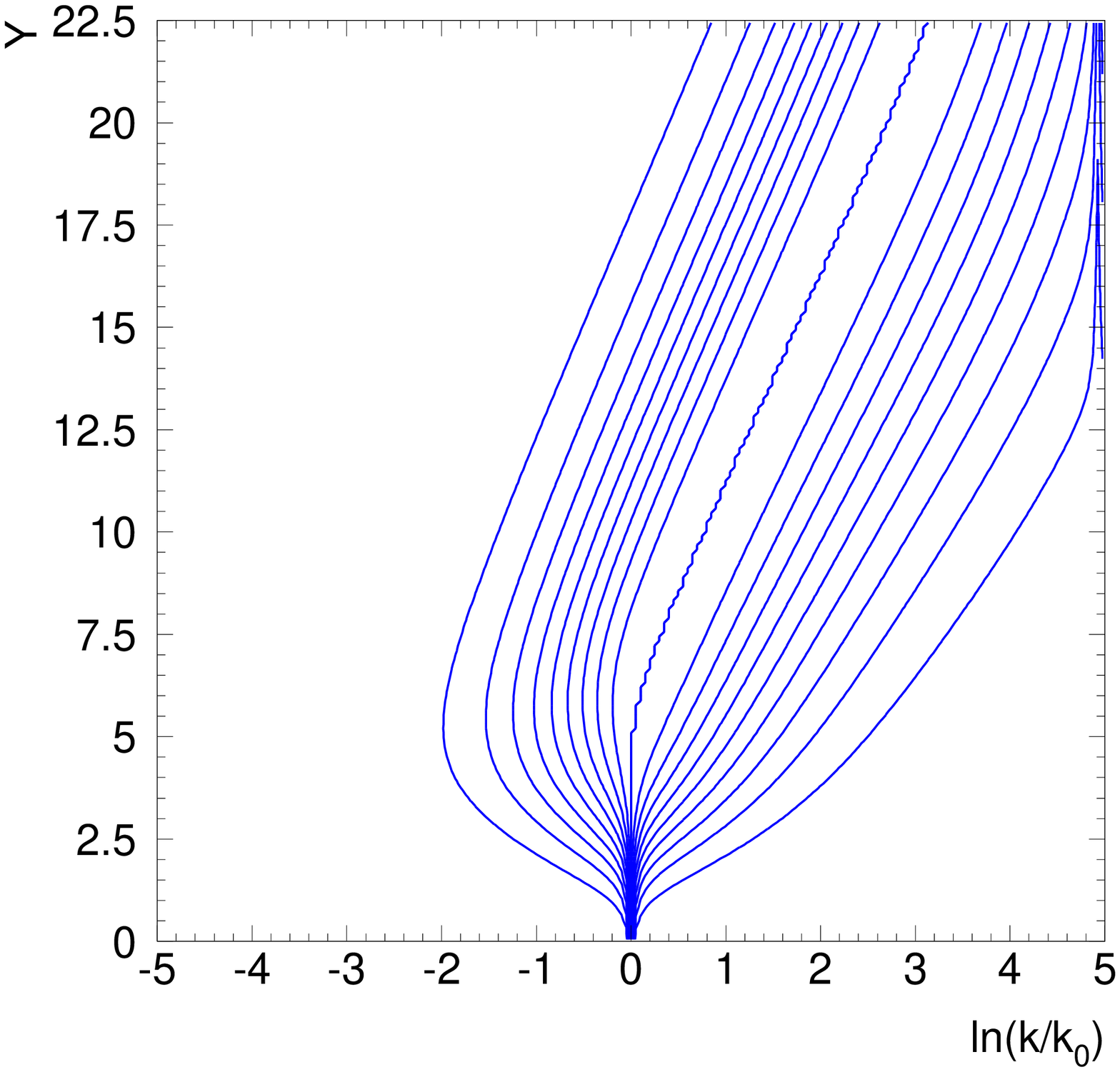,width=6.cm}
\caption{Contour plots of the renormalized distribution $\Psi(k,Y)$ in the case of the linear BFKL solution and the nonlinear BK solution.}
\label{fig:ContourBFKL}
\end{figure}

 The Gaussian shape  predicted by (\ref{eq:SaddleBFKL})  is obvious
and the  width increases with 
 rapidity. This consitutes a  potential  problem since while one  starts from a perturbative calculation  at a fixed, large scale $k_0$, eventually the  nonperturbative regime of $\Lambda_{QCD} \sim k \ll k_0$ is reached. 
This is not the case
for the solutions  nonlinear BK equation, which are illustrated by solid lines on right plot in  Fig.~\ref{fig:DiffBFKL}.
Clearly, in the case of the solutions to the BK equation,  the diffusion into the infrared region is strongly  suppressed. With increasing $Y$ the  distribution peak moves away from the initial value $k_0$
towards the  larger values of $k$.  Therefore  one can define saturation scale as a position of the peak
$$
Q_s(Y) \equiv k_{\rm max}(Y) \; .
$$
The suppression of the   diffusion  can be also visualized with help of the  normalized distribution
\be
\Psi(k,Y) \; = \; \frac{k\phi(k,Y)}{k_{\rm max}(Y)\phi(k_{\rm max}(Y),Y) } \; .
\label{eq:PsiNorm}
\ee
On left hand side of  Fig.~\ref{fig:ContourBFKL}  shows the contour plot in $(k,Y)$ space of ths distribution (\ref{eq:PsiNorm})
for the case of the linear BFKL equation. The contour lines correspond to the  constant values of the normalized distribution $\Psi(k,Y)$. The diffusive character of the solution  is clearly visible.
On the right hand side  of Fig.~\ref{fig:ContourBFKL}  we show the corresponding  contour plot
for  the nonlinear BK equation. We see that the contour lines are shifted towards the 
higher values of transverse momenta\footnote{A distortion of the contours at the highest values of $k$ is unphysical and is caused by cutoffs in numerical calculation.}. We can also identify a line in $(k,Y)$ space
which divides the  region where there  still is a diffusion (to the right)
and where there is no diffusion. The   contour lines are parallel to each other in the latter case what means that the solution is scaling there.
When the parametrisation $\xi=\ln k/k_0 -\lambda Y+\xi_0$ is introduced  the solution  depends only on $\xi$ alone. The critical line defines the saturation scale that was introduced in the previous
paragraphs. 
It turns out  that the nonlinear BK equation can be approximately treated as a linear diffusion equation with the absorptive boundary  close to the  critical line defined by the saturation scale $Q_s(Y)$ \cite{MuelTrian}.  This approximation allows us to evaluate precisely  the rapidity dependence of the saturation scale.

Recently, there has been quite substantial development achieved towards the understanding
of the solutions to the $(1+1)$ dimensional BK equation. In a  series of important papers \cite{MunPesch} it was proved that BK equation can be approximated as a diffusion equation with a nonlinear term. This makes it equivalent to the Fisher-Kolmogorov-Petrovsky-Piscounov
(FKPP) equation \cite{FKPP} 
\be
\partial_t u(t,x) = \partial_x^2(t,x)+u(t,x) [1-u(t,x)] \; ,
\label{eq:FKKP}
\ee
where the change of variables from $(Y,\ln k)$ to $(t,x)$ has been performed with simultaneous identification of $\phi\rightarrow u$.
FKPP equation  has been already studied in many fields of physics,  and its  solutions are very well understood. In particular it is well known that the FKPP equation has a travelling wave solution for large times (which are equivalent to large energies) which
is just a property of  geometrical scaling. For a  review see for example \cite{FKPPstudies}.

\subsection{BK equation with running coupling}

\begin{figure}
\centerline{\epsfig{file=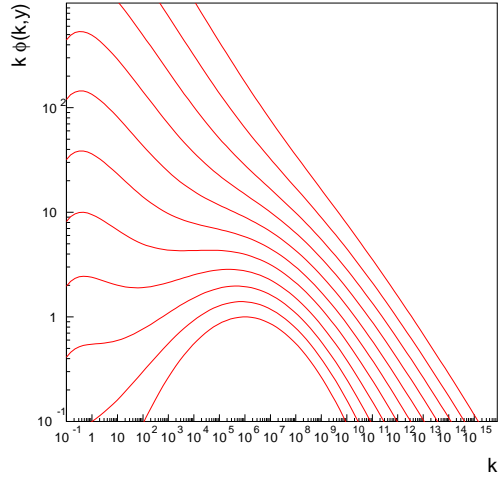,width=7.cm}}
\caption{Solution to the BFKL equation with running coupling.}
\label{fig:RunCoupling}
\end{figure}

As we have already stated, the BK equation has been derived within the leading logarithmic in $\ln 1/x $ (LLx) approximation in which
the coupling constant is fixed. However, it is very well known fact, that NLLx effects 
 are very important in the BFKL formalism \cite{NLLBFKL}. At NLLx order the coupling runs, and the linear BFKL equation becomes very unstable. The reason is  that the 
linear evolution is very sensitive to the details of the running coupling regularization.
This effect is illustrated in Fig.~\ref{fig:RunCoupling} where  we show the solution $k\phi^{\rm BFKL}(k,Y)$ with the running coupling, 
as a function of the transverse momentum for increasing values of $Y$. 
For small values of rapidity 
the position of the maximum  remains close to the value of the initial condition $k=k_0$.
However as the rapidity increases the position of the maximum abruptly shifts to the small
values of momenta.
The actual shape of the solution are critically dependent on the
 regularized value of the  coupling, $k\simeq k_{\rm reg} \ll k_0$.

With the inclusion of the running coupling the BK equation has the following form
\be
\frac{d \phi( k,Y)}{dY} \; = \; {\asb(k)} \int \frac{dk'}{k'} {{\cal K}(k,k')} \, \phi(k',Y) \, - \, {\asb(k)} \phi^2(k,Y)  \; .
\label{eq:BKRC}
\ee

\begin{figure}
\centerline{\epsfig{file=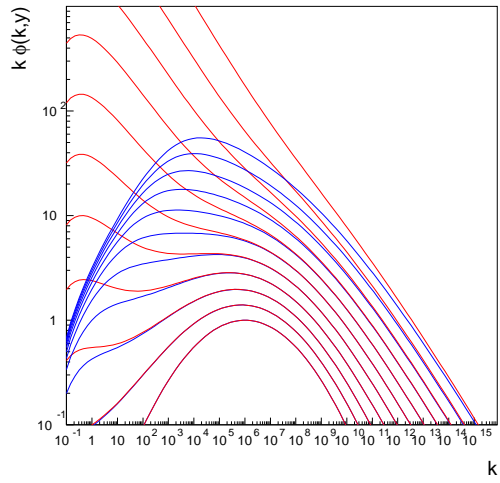,width=7.cm}}
\caption{Solution to the BK equation (as compared to BFKL) with running coupling .}
\label{fig:BKRC}
\end{figure}
The solution to the above equation is illustrated in Fig.~\ref{fig:BKRC},
 superimposed onto the solution of the linear BFKL equation.
The solution is much more stable than in the linear case, and
the maximum of the distribution  is not shifted to the infrared. Instead as rapidity increases, the maximum moves towards the higher values of transverse momenta. The reason for this is that the nonlinear term strongly damps the diffusion into  the infrared regime. The saturation scale $Q_s(Y)$ provides a natural cutoff for the low momenta and then no dependence on the regularisation of the running coupling is seen (see for example \cite{GBMS,BraunRC,AAMSW,DDD}).

One might ask whether the geometrical scaling is still preserved in the presence of the additional scale
$\Lambda_{QCD}$ which is implicitly introduced by use of the running coupling constant.
It turns out that the scaling  still holds, although the saturation scale has a  different rapidity dependence (see \cite{GBMS,MuelTrian,IIM,DDD,Dionisys} and also \cite{GLR})
\be
Q_s(Y) \; = \; \Lambda_{QCD} \exp\left(\sqrt{\frac{12 c}{\beta_0}(Y-Y_0)+\ln^2 Q_0/\Lambda_{QCD}}\right) \; ,
\label{eq:qsatRC}
\ee
with $c \simeq 2$.
The above formula has been derived by assuming that the local exponent of the saturation scale 
$$
\lambda(Y) \; = \; \frac{d\ln(Q_s(Y)/\Lambda)}{dY} \; 
$$
has the similar form as in the case of the fixed coupling 
$$
\lambda(Y) \; = \; c \alpha_s(Q^2_s(Y)) \; .
$$
By using these two formulae one can derive the saturation scale for the  running coupling   (\ref{eq:qsatRC}), see \cite{GBMS,IIM}.

\section{Solution in $(3+1)$ dimensions}
\subsection{Spatial distribution: impact parameter dependence}
\begin{figure}
\centerline{\epsfig{file=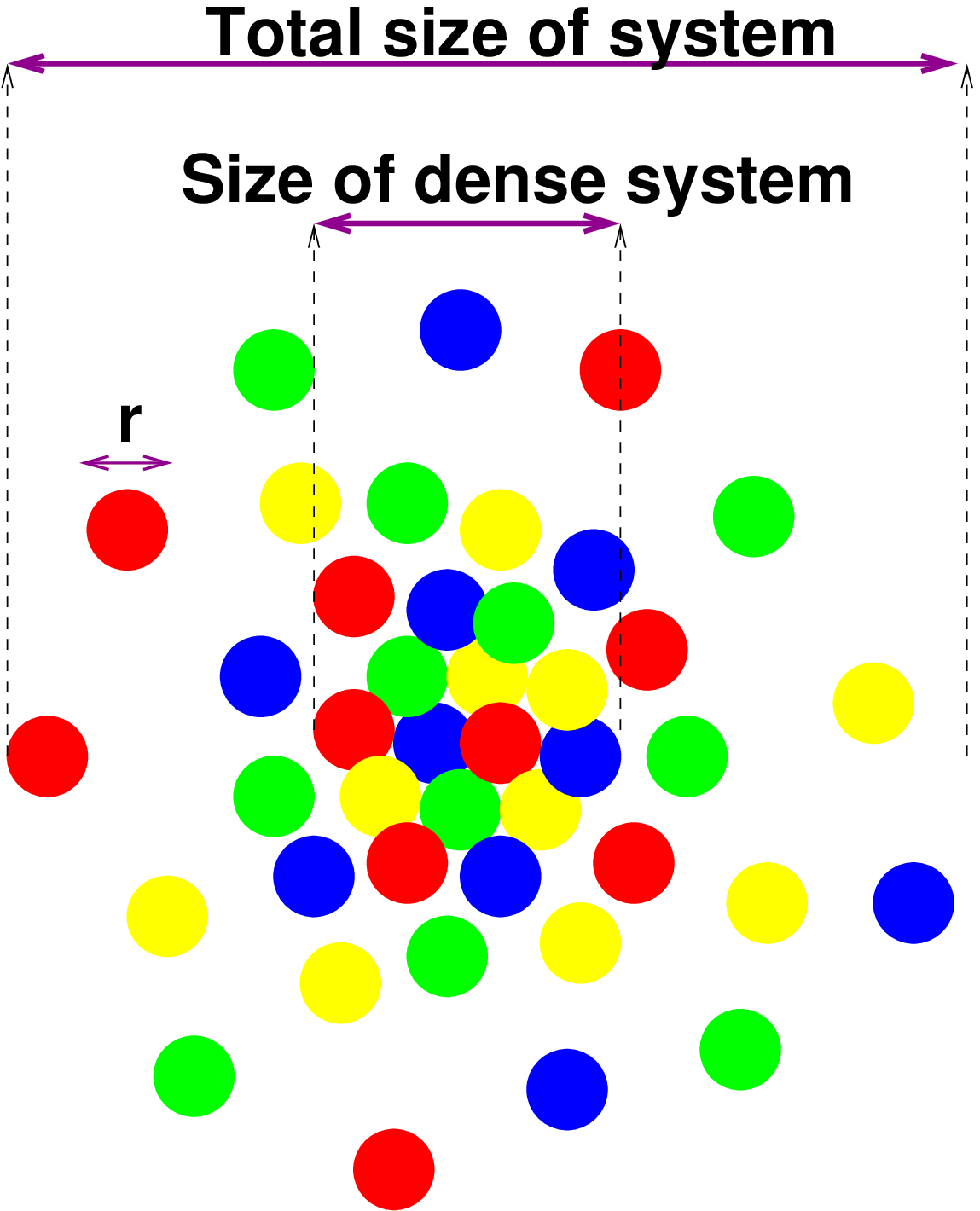,width=4.5cm}\hfill\epsfig{file=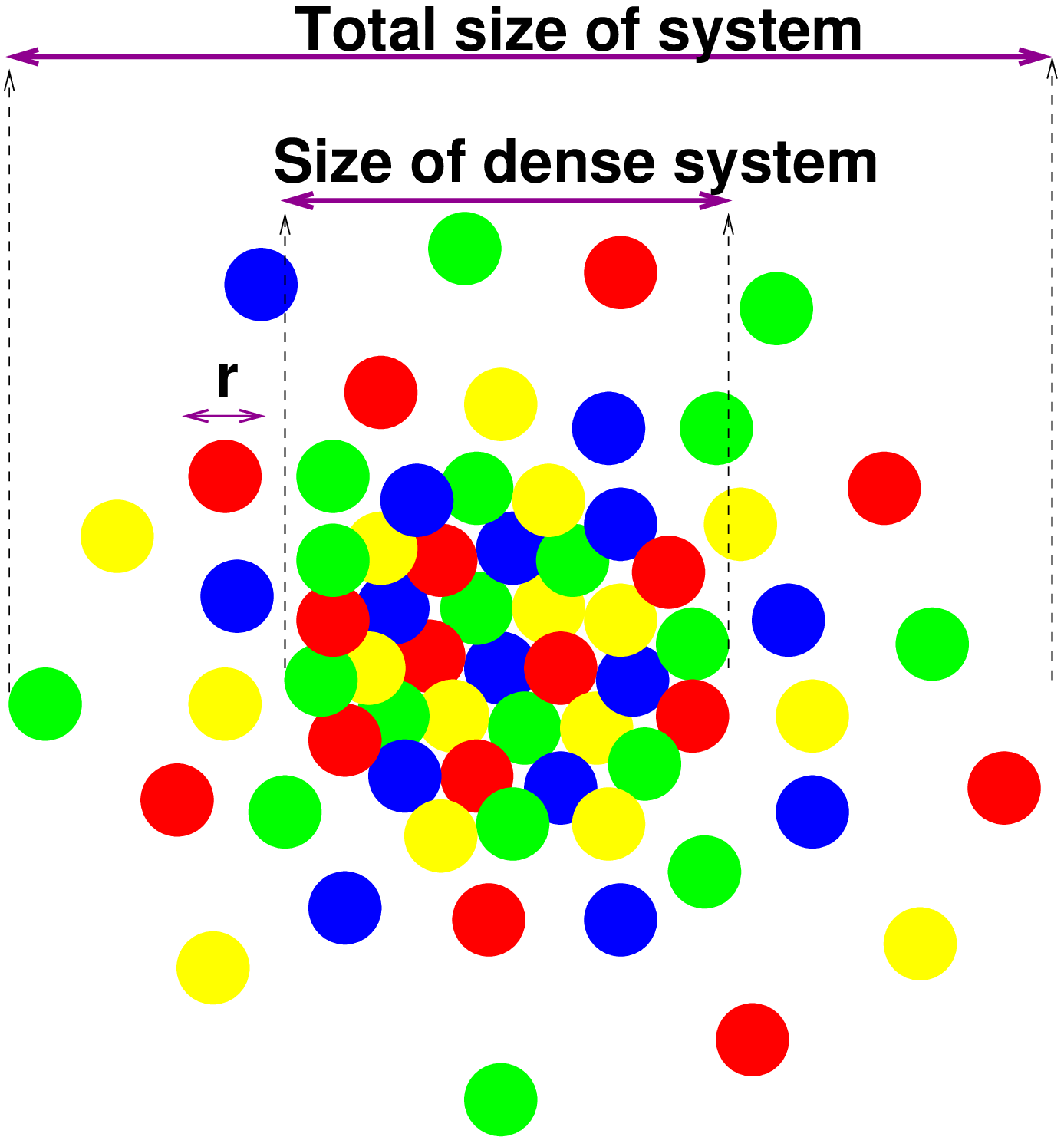,width=5.5cm}}
\caption{Schematic picture of the impact parameter dependence of the parton density at high energies. Right plot: as energy increases the area of the saturated region, the {\it black disc}, increases.}
\label{fig:IPpicture}
\end{figure}

\begin{figure}
\centerline{\epsfig{file=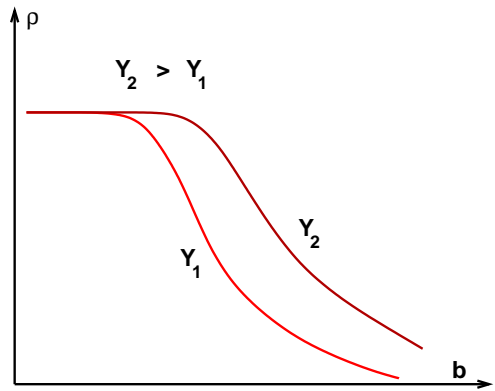,width=6.5cm}}
\caption{Impact parameter dependence of the scattering amplitude for two different values of rapidity. As energy (rapidity) increases the radius of the saturated plateau also grows.}
\label{fig:bdepsche}
\end{figure}

The phenomenon of saturation discussed so far has been based on the properties 
of the solutions to the BK equation in $1+1$ dimensions. Such approach  ignores the spatial distribution of the probe-target system. 
The solution to $(1+1)$ dimensional BK equation
 is integrated over the impact parameter. This is a result of the fact that in $1+1$
dimensional case we have assumed an infinite size of the target, and ignored any edge effects. One might expect that the more realistic picture of saturation looks as follows: as the probe collides with the target,  a dense system of partons emerges in the limited region of the impact parameter space.  For larger  impact parameters, the density of partons becomes more and more dilute.   The radius of the dense, saturated system expands with the growth of the energy.
This process is schematically drawn in Fig.~\ref{fig:IPpicture}.
One expects that the impact parameter profile of the scattering amplitude has the behaviour shown in Fig.~\ref{fig:bdepsche}.

\begin{figure}
\centerline{\epsfig{file=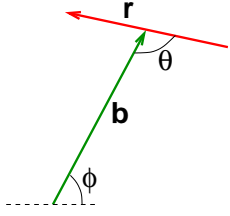,width=3.0cm}}
\caption{Parametrisation of dipole position.}
\label{fig:DipolePosition}
\end{figure}

A question thus arises  whether BK equation can provide  information about the impact parameter profile of the amplitude,  and whether this profile is consistent with the qualitative picture of saturation. 
Let us recall the BK equation with full dependence on all coordinates

\begin{multline}
 \frac{dN(\bzerone,\xzerone,Y)}{dY}  = 
  \asb \int   \frac{d^2 {\bf x}_2 \, \xzerone^2}{\xtwozer^2 \, \xonetwo^2} \,
 \bigg[ N(\bzerone+\frac{\xonetwo}{2},\xtwozer,Y)+ N(\bzerone-\frac{\xtwozer}{2},\xonetwo,Y) \,\\
 - \, N(\bzerone,\xzerone,Y)  \, - \,{ N(\bzerone+\frac{\xonetwo}{2},\xtwozer,Y)\,N(\bzerone-\frac{\xtwozer}{2},\xonetwo,Y)} \bigg] \;.
\label{eq:kov1}
\end{multline} 
As  stated before,
in general the problem is very difficult, even numerically, since one has four degrees of freedom per dipole plus rapidity as the evolution variable.
Even though the integral measure does not depend on the positions of the dipoles, the impact parameter dependence still exists. It is implicitly generated via the couplings of $b_{ij}$ to $x_{ij}$ in the 
arguments of the functions $N$, see Eq.~\ref{eq:kov1}.
 To simplify the problem, and yet retain the information about the impact parameter dependence, we note that measure in the equation  
$$
{ \frac{d^2 {\bf x}_2 \, \xzerone^2}{\xtwozer^2 \, \xonetwo^2}} \; ,
$$
is invariant under  global rotations in transverse space
$$
{ {\bf x}_0,{\bf x}_1,{\bf x}_2 \longrightarrow {\cal O}(\phi){\bf x}_0,{\cal O}(\phi){\bf x}_1,{\cal O}(\phi){\bf x}_2} \; ,
$$
see Fig.~\ref{fig:DipolePosition}.

Therefore one can assume that the position of the dipole is specified by three variables : the dipole size $r$, the impact parameter $b$, and the relative orientation of the dipole with respect to the impact parameter axis (angle $\theta$), see Fig.~\ref{fig:DipolePosition}.
 The invariance with respect to global rotations is equivalent to a condition  that
the target is cylindrically symmetrical.
Thus the problem reduces to $(3+1)$ dimensions: 
$N(r,b,\theta,\phi;Y)\; \rightarrow \; N(r,b,\theta;Y)$.
The BK equation with the impact parameter dependence has been investigated numerically \cite{GBS,LevinNaftali}, see also \cite{IkedaMcLerran}.
Here we show some of the results taken from \cite{GBS}, which used
the initial distribution in Glauber - Mueller form 
\be
N^{(0)}(r,b,\theta;Y=0) \; = \; 1-\exp(-r^2 S(b)) \;,
\label{eq:GMFormula}
\ee
where the impact parameter profile has been chosen to be of the Gaussian type
\be
S(b) \; = \; \frac{1}{R_0^2} \exp(-b^2 /b_0^2) \; .
\ee

\subsection{Impact parameter dependence}
\begin{figure}
\centerline{\epsfig{file=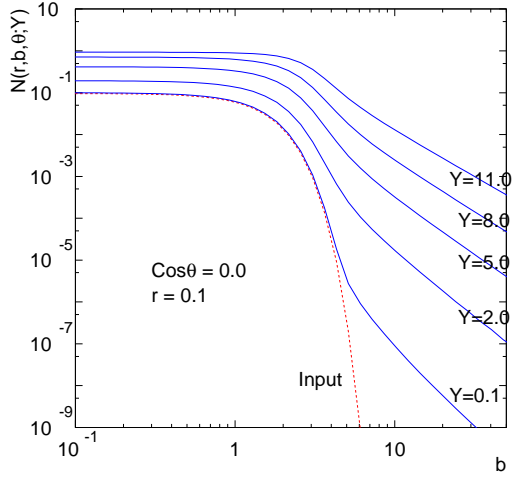,width=7cm}}
\caption{Impact parameter dependence of the solution to the BK equation for increasing values of rapidity.}
\label{fig:bdepkov}
\end{figure}

\begin{figure}
\centerline{\epsfig{file=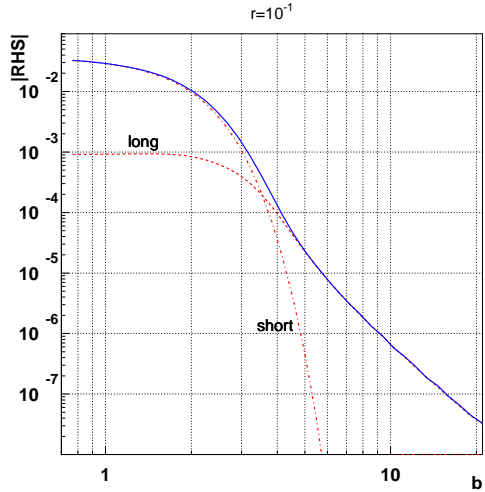,width=7cm}}
\caption{The impact parameter dependence of the amplitude decomposed into the short and long range contributions after small evolution step in rapidity $\Delta Y=0.1$. }
\label{fig:shortlong}
\end{figure}

In Fig.~\ref{fig:bdepkov} we show the resulting impact parameter dependence of the BK solutions calculated for different values of rapidity $Y$. For small values of $b$  the amplitude is large and strong nonlinear effects are evident. On the other hand, for large values of $b$ one observes a fast growth of
the amplitude is governed by the linear part of the equation. One can verify that the increase is exponential in rapidity as expected from  the BFKL equation.
 The region in impact parameter space, where the amplitude is large, expands  with growing rapidity. However, perhaps the most striking feature  is the fact that the initial profile in impact parameter is not preserved even after a small step in rapidity, $\Delta Y = 0.1$. The exponential tail of the initial distribution  immediately assumes power behaviour $\sim 1/b^4$.
It is interesting to investigate the origin of these power tails. 
One can divide the region of integration into  long- and short-range parts   by introducing a separating cutoff $r_0$ on the dipole size.

\begin{multline}
\left[\stackrel{short}{\overbrace{\int { \Theta(r_0-|\xbtwo-\bb|)}}}+\stackrel{long}{\overbrace{\int { \Theta(|\xbtwo-\bb|-r_0)}}} \right] \frac{d^2 \xbtwo (\xbzero-\xbone)^2}{(\xbzero-\xbtwo)^2 (\xbone-\xbtwo)^2} \nonumber \\
 \cdot \left(N_{02}^{(0)}+N_{12}^{(0)}-N_{01}^{(0)}-N_{02}^{(0)} N_{12}^{(0)} \right)\; . \nonumber
\end{multline}
In Fig.~\ref{fig:shortlong} the profile in impact parameter space has been decomposed into short and long range contributions. One can see that the short range contribution dominates  the behaviour at small values of $b$.
There the exponential behaviour is preserved since we have the factorisation of   the initial profile $S(b)$ at small values of $b$. The long range contribution is dominating at large values of impact parameter $b$ where it generates the power tail. Thus the $\sim 1/b^4$ behaviour originates from the integration of the large dipole sizes and is a reflection of the asymptotic behaviour of the integral kernel (see discussion in \cite{KovnerWiedemann}).

\subsection{Violation of Froissart bound}

The presence of the power tail in impact parameter has profound consequences for the unitarity. Power decrease of the amplitude means that the interaction is long range. As already stated it, is a direct consequence of the power-like form of the integral kernel
$$
{  \frac{d^2 {\bf x}_2 \, \xzerone^2}{\xtwozer^2 \, \xonetwo^2} \simeq
  d^2 {\bf x}_2 \, \frac{r^2}{b^4}} \; .
$$
This type of fast expansion of the interaction system leads to the violation of the Froissart bound, as has been first observed in \cite{KovnerWiedemann} (compare also a parallel discussion in \cite{FIIM}). It turns out that even though the amplitude is equal or less than $1$, due to the nonlinearity of the equation, the dipole cross section
increases fast with the decreasing $x$ which violates Froissart bound
$$
\sigma \; =\; \int d^2{\bf b} \,  N({\bf r},{\bf b}; Y=\ln 1/x) \;  \sim  \; x^{-\lambda} \; .
$$
This happens because the kernel in the  BK equation is conformally invariant, with no mass scale which would cut off the long range contributions.

\subsection{Dipole  size dependence}
\begin{figure}
\centerline{\epsfig{file=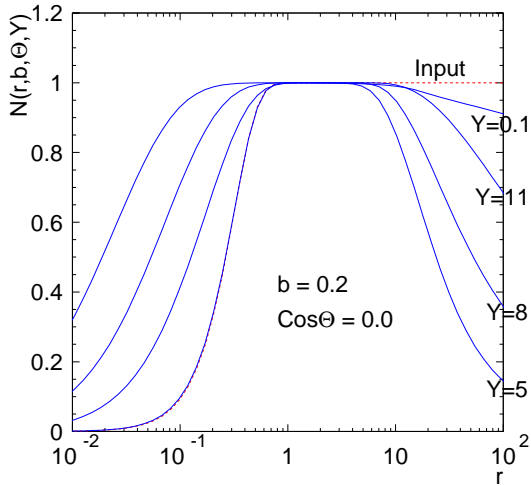,width=7cm}}
\caption{Solution of the impact parameter dependent BK equation with fixed orientation and position of the dipole for various values of rapidity $Y=0.1,5,8,11$.}
\label{fig:rdepbsmall}
\end{figure}

\begin{figure}
\centerline{\epsfig{file=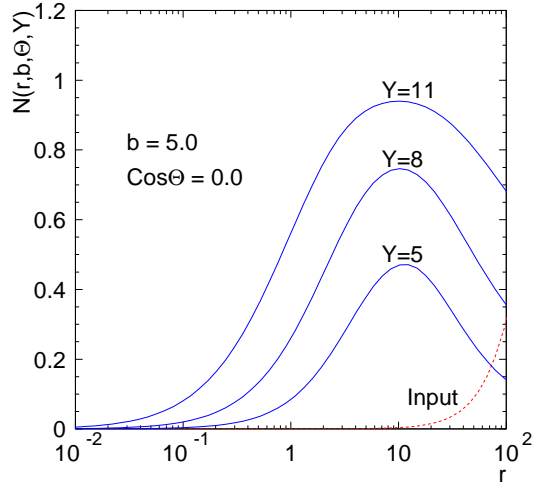,width=7cm}}
\caption{The same as Fig.~\ref{fig:rdepbsmall} but for large value of impact parameter $b=5$.}
\label{fig:rdepblarge}
\end{figure}

The dipole size dependence in the case of the impact parameter dependent BK equation
is shown in Fig.~\ref{fig:rdepbsmall} where the value for the impact parameter has been chosen to correspond to a  central collision. For small and moderate values of $r$ it has qualitatively the same behaviour as previously,  the amplitude vanishes as $r$ tends to $0$ and extends to lower values of $r$ as rapidity grows. It also saturates to $1$ for moderate values of $r$. However,  at large values of the dipole size the situation is dramatically different. Here the amplitude drops down again. The reason is that now there is  a dimension in impact parameter which characterizes the size of the target. As the dipole grows, at some point it has to completely miss the target and amplitude becomes zero again. This is quite different from the previous case (without the impact parameter) where the amplitude was always saturated since there was an  infinite target. 
It is also interesting to study the solutions at  different values of $b$. In Fig.~\ref{fig:rdepblarge} we present the dipole size dependence of the amplitude for larger value of $b$ which corresponds to a peripheral collision of the dipole
with the target. In that case we see that the amplitude peaks for values of the dipole size twice the size of the  impact parameter. This is expected, since it reflects the 
properties of the integral kernel in the  BK equation.
The solution in the b-dependent case has also several other interesting properties.
For values of dipole size much smaller than  the impact parameter $r \ll b$, the amplitude depends only on a single combined variable $r^2/b^4$. This dependence on one variable, the {\it anharmonic ratio }, is the result of the conformal symmetry, see for example \cite{Lipatov86}.

\subsection{Saturation scale with $b$ dependence}
One can also extract the saturation scale from the solution to BK equation in (3+1) dimensions.  In \cite{GBS} a following prescription has been used
\be
\langle N(r=1/Q_s,b,\theta;Y)\rangle_{\theta} \; = \; \kappa, \; \kappa\sim 0.5 \; .
\label{eq:bdepSS}
\ee
From Fig.~\ref{fig:rdepbsmall} we see that the above equation possesses two solutions
$$
\frac{1}{Q_s(b,Y)} < r < R_H(b,Y) \; .
$$

\begin{figure}
\centerline{\epsfig{file=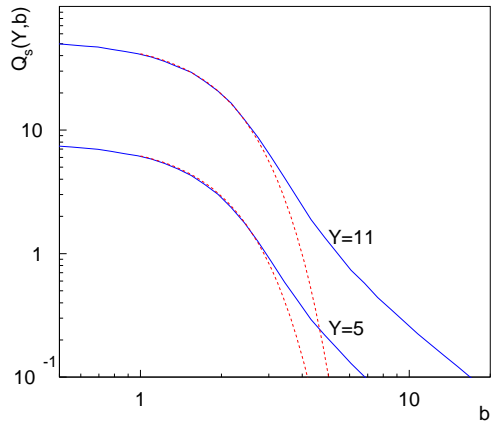,width=7cm}}
\caption{Impact parameter profile of the saturation scale for two various rapidities $Y=5$ and $Y=11$. Solid lines are the result of the calculation of full BK equation. Dashed lines correspond to the exponential behaviour of the saturation scale that was expected from initial conditions.}
\label{fig:bdepSS}
\end{figure}

The lower bound $Q_s(b,Y)$ is the impact parameter dependent saturation scale, which has been plotted in Fig.~\ref{fig:bdepSS}. We see that the saturation scale has strong dependence on $b$, it is largest at small values of impact parameter and then decreases for large values of impact parameter. The physical picture is that while the impact parameter is increased, one moves from a strongly saturated regime to a more dilute one.
The tail of the saturation scale is again power like $\sim 1/b^2$ which is to be expected from the properties of the integral kernel. We see that this behaviour is  different from the one that could be anticipated from initial conditions (compare dashed lines).
The saturation scale has following behaviour
$$
Q_s^2(b,Y) \simeq g(b) \, \exp(\asb 2 \lambda_s Y), \; \lambda_s \simeq 2 \; ,
$$
where function $g(b)$ is exponentially falling for small values of $b$ and has a power like behaviour at large values of impact parameter.

The second solution $R(b,Y)$ is  new compared to the b-independent case, and it just reflects the fact the there is an additional scale present, the finite size of the target.

\section{Conclusions and outlook. Beyond BK equation}

We have described basic properties of the BK equation which is a nonlinear evolution equation suitable for the description of partonic systems at high density. 
We  have shown that the solution to this equation has a property of geometrical scaling 
with the characteristic saturation scale.  The nice property of this equation is the suppression of the infrared diffusion and independence of the regularization for the running coupling. 
The impact parameter dependence of this equation leads to the violation of the Froissart bound despite the fact that the amplitude is bounded from above. This is a consequence of purely perturbative approach and the lack of long distance effects such as confinement in the BK equation.

However, the  BK equation has been derived by using strong assumptions about lack
of correlations in the system and it is thus an equation in mean field approximation. 
It is not clear to what extent the BK equation is a good approximation to the full Balitsky-JIMWLK equations.
In numerical studies of the dipole scattering by Salam \cite{Salam} it has been shown that
the fluctuations are very important  and lead to a very different result as compared with the 
mean field approach.  
A lot of theoretical effort has been recently devoted to study the role of correlations:
in \cite{Iancu-Mueller} a more quantitative study of the fluctuations has been proposed, in \cite{Levin-Lublinsky} a new equation for the generating functional was postulated which takes into account correlations in the nuclei; in \cite{Mueller-Shoshi} a BK equation with two absorptive boundaries has been studied; in \cite{Weigert-Rummukainen} a numerical study of the full JIMWLK equation has been performed for the first time; in \cite{Peschanski-Janik}  an analytical study of the Balitsky hierarchy 
restricted to the dipole operators, and in \cite{IMM,IT} the role of the discreteness of the gluon system
 and connection to the statistical physics have been discussed.
We expect that this line of research will be continued in the near future  and
we will be able to understand the fascinating and complex theory of strong interactions even better.

\section*{Acknowledgments}
This research is supported by U.S. Department of Energy, Contract No. DE-AC02-98CH10886. I thank Peter Rembiesa for critically reading the manuscript.

\end{document}